\documentclass[conference]{IEEEtran}
\usepackage{cite}
\usepackage{amsmath,amssymb,amsfonts}
\usepackage{algorithmic}
\usepackage{graphicx}
\usepackage{textcomp}
\usepackage{fancyhdr}
\usepackage[hyphens]{url}
\usepackage{braket}
\usepackage[table,xcdraw,dvipsnames]{xcolor}

\def\BibTeX{{\rm B\kern-.05em{\sc i\kern-.025em b}\kern-.08em
    T\kern-.1667em\lower.7ex\hbox{E}\kern-.125emX}}
    
\begin{document}
\title{NISQ+: Boosting quantum computing power by approximating quantum error correction}

\author{\IEEEauthorblockN{
Adam Holmes\IEEEauthorrefmark{2}\IEEEauthorrefmark{3},
Mohammad Reza Jokar\IEEEauthorrefmark{3},
Ghasem Pasandi\IEEEauthorrefmark{4},
Yongshan Ding\IEEEauthorrefmark{3},
Massoud Pedram\IEEEauthorrefmark{4},
Frederic T. Chong\IEEEauthorrefmark{3}}
\\[1em]
\begin{tabular}{*{3}{>{\centering}p{.30\textwidth}}}
\IEEEauthorrefmark{2}\textit{\small{Intel Labs}}&
\IEEEauthorrefmark{3}\textit{\small{Department of Computer Science}}&
\IEEEauthorrefmark{4}\textit{\small{Viterbi School of Engineering}}\tabularnewline
\small{Intel Corporation} &
\small{The University of Chicago} &
\small{The University of Southern California}\tabularnewline
\small{Hillsboro, OR 97214, USA}&
\small{Chicago, IL 60637, USA}&
\small{Los Angeles, CA 90007, USA}\tabularnewline
\small{adam.holmes@intel.com} & &
\end{tabular}}
%
\maketitle


\maketitle
\pagestyle{plain}


\begin{abstract}

Quantum computers are growing in size, and design decisions are being made now that attempt to squeeze more computation out of these machines. In this spirit, we design a method to boost the computational power of near-term quantum computers by adapting protocols used in quantum error correction to implement ``Approximate Quantum Error Correction (AQEC)." By approximating fully-fledged error correction mechanisms, 
we can increase the compute volume (qubits $\times$ gates, or ``Simple Quantum Volume (SQV)") of near-term machines.
The crux of our design is a fast hardware decoder that can approximately decode detected error syndromes rapidly. Specifically, we demonstrate a proof-of-concept that approximate error decoding can be accomplished online in near-term quantum systems by designing and implementing a novel algorithm in Single-Flux Quantum (SFQ) superconducting logic technology.
This avoids a critical decoding backlog, hidden in all offline decoding schemes, that leads to idle time exponential in the number of T gates in a program \cite{terhal2015quantum}.

Our design utilizes one SFQ processing module per physical quantum bit.  Employing state-of-the-art SFQ synthesis tools, we show that the circuit area, power, and latency are within the constraints of typical, contemporary quantum system designs.
Under a pure dephasing error model, the proposed accelerator and AQEC solution is able to expand SQV by factors between 3,402 and 11,163 on expected near-term machines. The decoder achieves a $5\%$ accuracy threshold as well as pseudo-thresholds of approximately $5\%, 4.75\%, 4.5\%,$ and $3.5\%$ physical error rates for code distances $3, 5, 7,$ and $9$, respectively. Decoding solutions are achieved in a maximum of $\sim 20$ nanoseconds on the largest code distances studied. By avoiding the exponential idle time in offline decoders, we achieve a $10$x reduction in required code distances to achieve the same logical performance as alternative designs.

\end{abstract}

\section{Introduction}\label{sec:Introduction}

\begin{figure}[t!]
  \centering
  \includegraphics[width=0.8\linewidth]{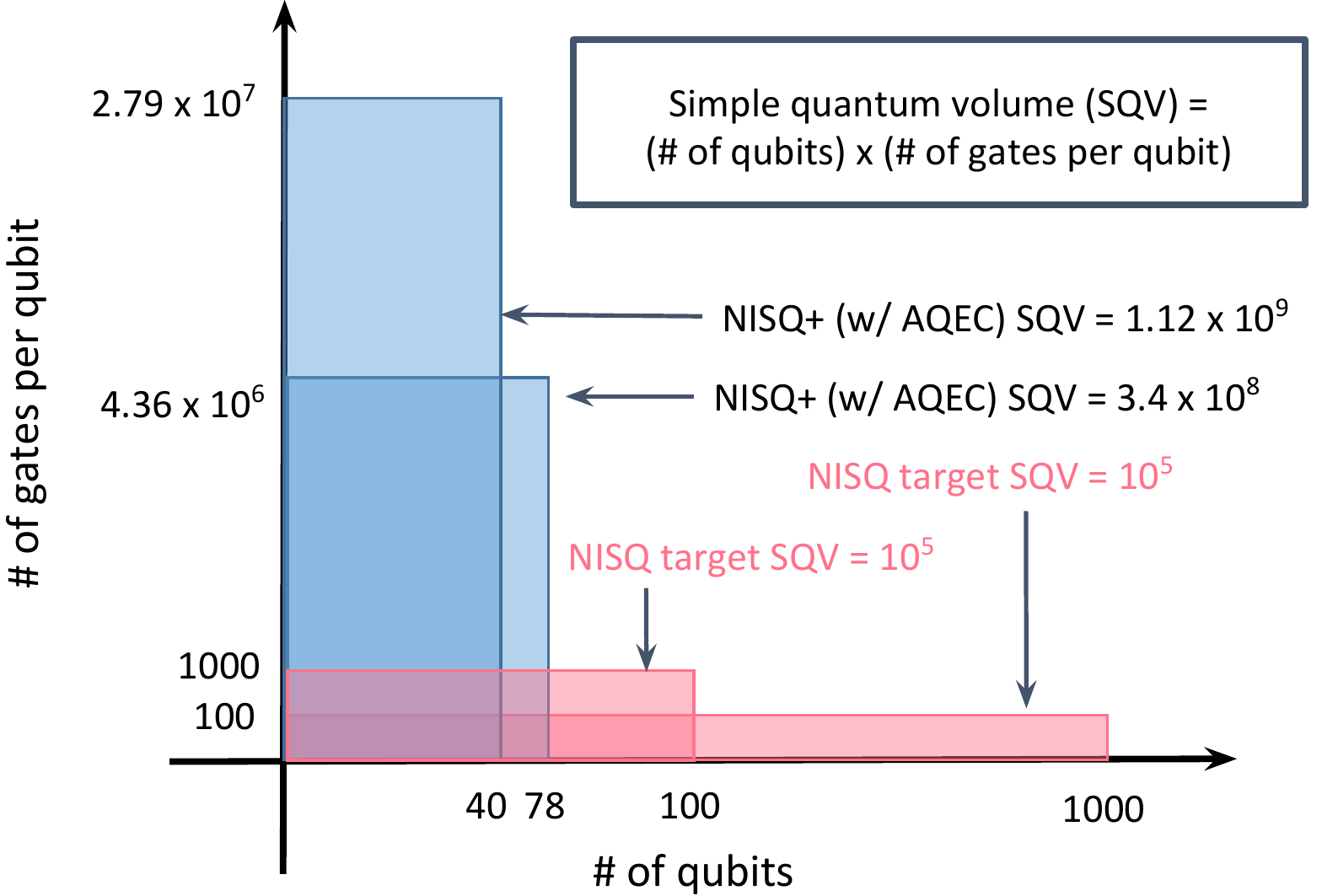}
  \caption{Boosting the quantum computation power with approximate error correction schemes. A machine with $1,024$ faulty physical qubits of error rate $10^{-5}$ has an SQV of $\approx 10^8$. By performing fast, online, approximate decoding, we can trade the number of computational qubits for gate fidelity and boost the SQV by over a factor of 3,402. Moving to a higher code distance raises this increase to a factor of 11,163. NISQ machines are severely limited by gate fidelity, and introducing error mitigation techniques can have dramatic effects on SQV.}
  \label{fig:sqv}
\vspace{-0.7cm}
\end{figure}

Quantum computing has the potential to revolutionize computing and have massive effects on major industries including agriculture, energy, and materials science by solving computational problems that are intractable with conventional machines~\cite{hastings2014improving,svore2016quantum}. As we begin to build quantum computing machines of between 50 and 100 qubits \cite{preskill2018quantum} and even larger (e.g. $1,000$), we are making design decisions to attempt to get the most computation out of a machine, or expand the "Simple Quantum Volume" (SQV). SQV can be defined as the number of computational qubits of a machine multiplied by the number of gates we expect to be able to perform without error, as in Figure~\ref{fig:sqv}. One limiting factor on SQV now is that physical quantum bits (qubits) are extremely error-prone, which means that computation on these machines is bottlenecked by the short lifetimes of qubits. System designers combat this by attempting to build better physical qubits, but this effort is extremely difficult and classical systems can be used to alleviate the burden. Specifically, $\emph{quantum error correction}$ is a classical control technique that decreases the rate of errors in qubits and expands the SQV. Error correction proceeds by encoding a set of \emph{logical} qubits to be used for algorithms into a set of faulty physical qubits. Information about the current state of the device, called \emph{syndromes}, is extracted by a specific quantum circuit that does not disturb the underlying computation. Decoding is the process by which an error correcting protocol maps this information to a set of \emph{corrections} that, if chosen correctly, should return the system to the correct logical state. Fully fault tolerant machines can expand the SQV rapidly by suppressing qubit errors exponentially with the code distance.



While a fully fault-tolerant quantum computer may take many years to construct, it is possible to use the well-developed theory of error correction as inspiration for constructing $\emph{error mitigation}$ protocols that still provide a strong expansion in SQV. In this paper we present an approximate decoding solution specifically targeting execution time and show that we can in fact perform decoding at the speed of syndrome generation for near-term machines. Prior work has suggested and analyzed software solutions for decoding, but relying on hardware-software communication can be slow, especially considering the cryogenic environment of typical quantum computing systems. If decoding occurs slower than error information is generated, the system will generate a backlog of information as it waits for decoding to complete, introducing an exponential time overhead that will kill any quantum advantage (see Section \ref{sec:fast}). A hardware solution proposed here results in the ability to perform logical gates with orders of magnitude better fidelity and at the speed of syndrome generation, resulting in a major expansion in SQV as shown in Figure~\ref{fig:sqv}. This relies on an approximate decoding algorithm implemented in superconducting Single Flux Quantum (SFQ) hardware. While the algorithmic design enables the accuracy of the hardware accelerator to be competitive at small scale with existing software implementations, the benefits of implementing the circuitry directly in SFQ hardware are numerous. Specifically, high clock speeds, low power dissipation, and unique gating style allows for our accelerator to be co-located with a quantum chip inside a dilution refrigerator, avoiding otherwise high communication costs. 

This work contributes the following:
\begin{enumerate}
    \item We design the first approximate decoding algorithm for stabilizer codes based on SFQ hardware, leveraging unique capabilities that the hardware offers,
    \item We show that using this new error mitigation technique, we can expand the SQV of near-term machines by factors of between 3,402 and 11,163,
    \item We use Monte-Carlo simulation based benchmarking of the hardware accelerator, resulting in effective accuracy and pseudo-thresholds,
    \item We perform system execution time analysis, realistically benchmarking the decoder performance in real time and showing that decoding is likely to be able to proceed at or exceeding the speed of data generation enabling the benefits of fault tolerant quantum computing.
    \item We show that our online decoder requires $10x$ smaller code distance than offline decoders when decoding backlog accounted for.  
\end{enumerate}

The remainder of the paper is as follows: Section \ref{sec:Background} describes the necessary background of quantum computation and details the specifications of typical quantum computing systems stacks. Section \ref{sec:quantumerrorcorrection} describes quantum error correction and the decoding problem in detail. Section \ref{sec:related} describes relevant related work in the area ranging from optimized software implementations of matching algorithms to novel descriptions of neural network based decoders. Section \ref{sec:decoder} describes our decoding algorithm, and Section \ref{sec:implementation} describes implementation details of SFQ technology, and the circuit datapaths in detail. Section \ref{sec:methodology} describes our methodology for evaluation, including details of the simulation environment in which our accelerator was benchmarked, details of the metrics used to evaluate performance, and descriptions of novel synthesis tools used to generate efficient layouts of SFQ circuitry. Section \ref{sec:evaluation} presents our accuracy results, a breakdown of the accelerator characterization including area, power, and latency footprints, a timing evaluation, and analysis of the SQV effects. Section \ref{sec:conclusion} concludes.

\section{Background}\label{sec:Background}
In this section we discuss the basics of quantum computation, quantum error correction, and a description of the fundamental components of a quantum computing system architecture.
\subsection{Basics of Quantum Computation}
Here we provide a brief overview of quantum computation necessary to discuss quantum error correction. For more detailed discussions see \cite{MikenIke}. A quantum computing algorithm is a series of operations on two level quantum states called \emph{qubits}, which are quantum analogues to classical bits. A qubit state can be written mathematically as a superposition of two states as $\ket{\psi} = \alpha \ket{0} + \beta \ket{1}$, where the coefficients $\alpha, \beta \in \mathbb{C}$ and $|\alpha|^2 + |\beta|^2 = 1$. A measured qubit will yield a value of $\ket{0}$ or $\ket{1}$ with probability $|\alpha|^2$ or $|\beta|^2$, respectively, at which point the qubit state will be exactly $\ket{0}$ or $\ket{1}$. Larger quantum systems are represented simply as $\ket{\psi} = \sum_i \alpha_i \ket{i}$ where $\ket{i}$ are computational basis states of the larger quantum system. 

Quantum operations (gates) transform qubit states to other qubit states. In this work we will be making use of particular quantum operations known as \emph{Pauli gates}, denoted as $\{I, X, Y, Z\}$. These operations form a basis for all quantum operations that can occur on a single qubit, and therefore any operation can be written as a linear combination of these gates. 
Additionally, error correction circuits make use of the Hadamard gate $H$, an operation that constructs an evenly weighted superposition of basis elements when acting on a basis element. 
Two-qubit controlled operations will also be used, 
which can generate entanglement between qubits and are required to perform universal computation. 

\subsection{Quantum Error Correction}
\label{sec:quantumerrorcorrection}
Qubits are intrinsically fragile quantum systems that require isolation from environmental interactions in order to preserve their values. \emph{Decoherence}, for example the decay of a quantum state from a general state $\ket{\psi} = \alpha\ket{0} + \beta\ket{1}$ to the ground state $\ket{\psi'} = \ket{0}$ happens rapidly in many physical qubit types, often on the order of tens of nanoseconds \cite{tomita2014low,tannu2017taming}. This places a major constraint on algorithms: without any modifications to the system, algorithms can only run for a small, finite time frame with high probability of success.

To combat this, \emph{quantum error correction} protocols have been developed. These consist of encoding a small number of \emph{logical} qubits used for computation in algorithms into a larger number of physical qubits, resulting in a higher degree of reliability \cite{lidar2013quantum, dennis2002topological, fowler2012surface, terhal2015quantum}. In general, developing quantum error correction protocols is difficult as directly measuring the qubits that comprise a system will result in destruction of the data. To avoid this, protocols rely upon indirectly gathering error information via the introduction of extra qubits that interact with the primary set of qubits and are measured. This measurement data is then used to infer the locations of erroneous data qubits. 

While many different types of protocols have been developed, this work focuses primarily on the \emph{surface code}, a topological stabilizer code \cite{gottesman1997stabilizer} that is widely considered to be the best performing code for the medium-term as it relies purely on geometrically local interactions between physical qubits greatly facilitating its fabrication in hardware, and has been shown to have very high reliability overall \cite{fowler2012surface}.
\subsection{The Surface Code}


\begin{figure}[t!]
  \centering
  \includegraphics[width=\linewidth]{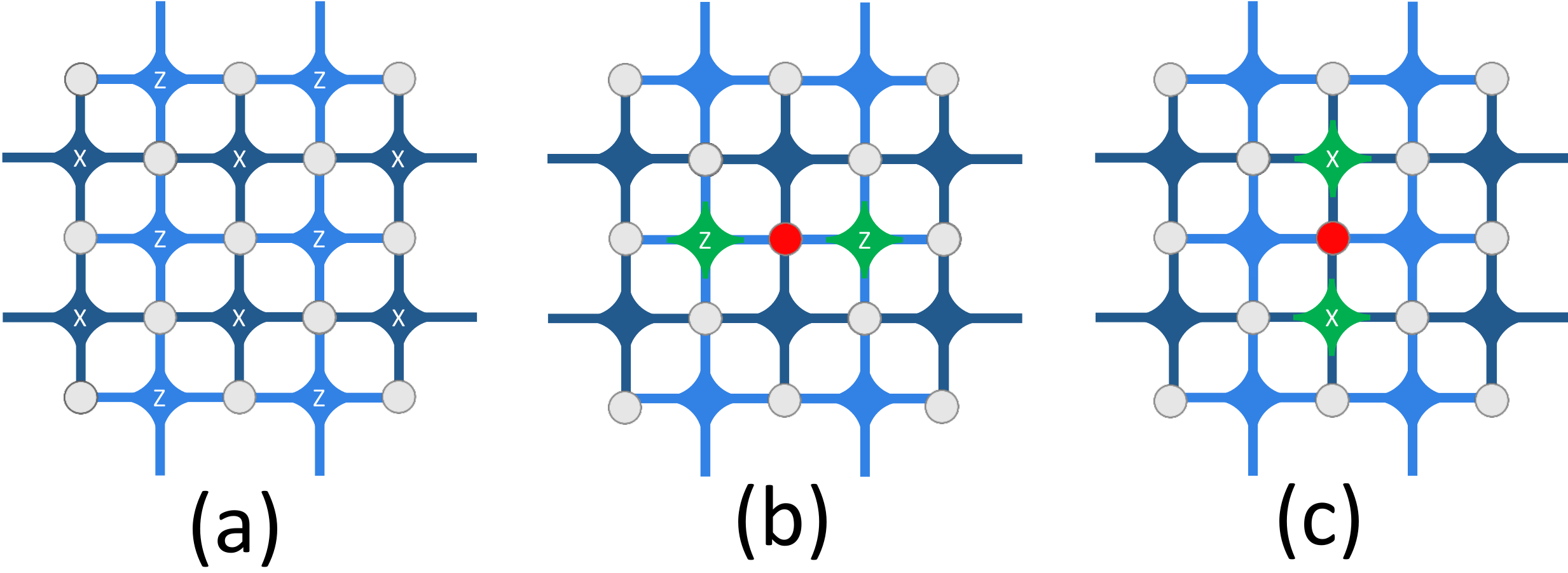}
  \caption{Figure (a) shows a graphical illustration of a surface code mesh. Gray circles indicate data qubits, and nodes labeled $X$ and $Z$ indicate ancillary qubits measuring $X$ and $Z$ stabilizers, respectively. Ancillary qubits are joined by colored edges to the data qubits that they are responsible for measuring. In figure (b) a single data qubit experiences a Pauli $X$ error indicated by red coloring, causing the neighboring $Z$ ancillary qubits to detect an odd parity in their data qubit sets and return $+1$ measurement values indicated by green coloring. In figure (c), the data qubit in red experiences a Pauli $Z$ error, causing the vertically adjacent $X$ ancillary qubits to return $+1$ measurement values. The entire error syndrome strings for either of these two cases would include a string of 12 values, two of which would be $+1$ and the remaining 10 would be $0$.}
  \label{fig:surface_mesh}
\vspace{-0.5cm}
\end{figure}
Errors can occur on physical qubits in a continuous fashion, as each physical qubit is represented mathematically by two complex coefficients that can change values in a continuous range. However, a characteristic of the quantum mechanics leveraged by the surface code is that these continuous errors can be \emph{discretized} into a small set of distinct errors. In particular, the action of the surface code maps these continuous errors into Pauli error operators of the form $\{I, X, Y, Z\}$ occurring on the data. This is one of the main features of the code that allows error detection and correction to proceed. 

\begin{figure}[t!]
  \centering
  \includegraphics[width=\linewidth]{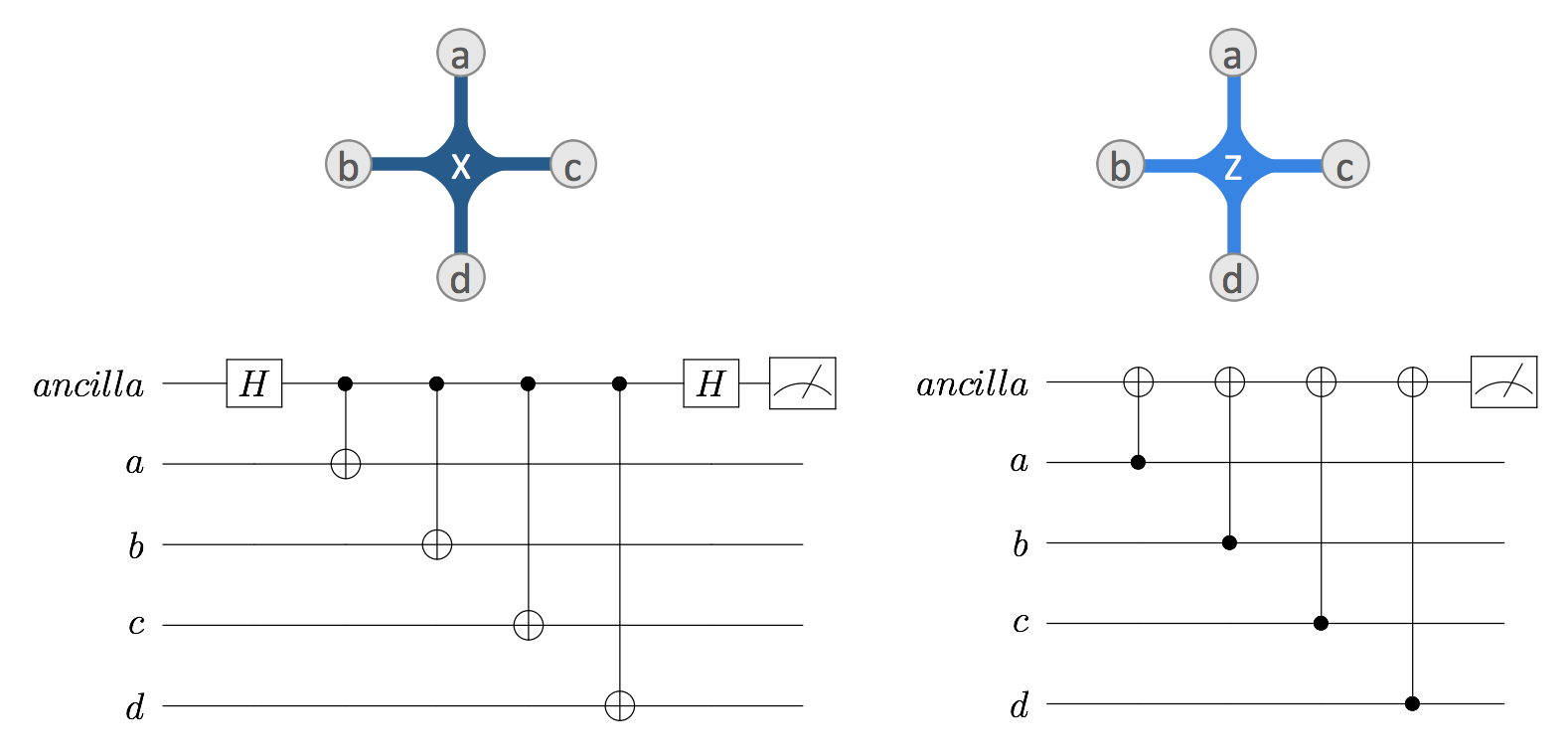}
  \caption{The ``X" (left) and ``Z" (right) stabilizer circuits required to generate error syndrome information. Horizontal lines indicate physical qubits, and boxes on these lines indicate single qubit operations. Vertical lines connecting qubits indicate two qubit operations, with solid circles signifying control qubit operands and open circles signifying controlled Pauli $X$ operations on target qubits. The ancilla qubits end the circuits with a measurement operator, signified by the final box.}
  \label{fig:surface_stabilizers}
\vspace{-0.6cm}
\end{figure}

The surface code procedure that accomplishes error discretization, detection, and correction is an error correcting code that operates upon a two-dimensional lattice of physical qubits. The code designates a subset of the qubits as data qubits responsible for forming the logical qubit, and others as ancillary qubits responsible for detecting the presence of errors in the data. This is shown graphically in Figure \ref{fig:surface_mesh}. Ancillary qubits interact with all of their neighboring data qubits and are then measured, and the measurement outcomes form the \emph{error syndrome}. This set of operations forms the \emph{stabilizer circuit}, where each ancillary qubit measures a four-qubit operator called a \emph{stabilizer} as in Figure~\ref{fig:surface_stabilizers}.


\subsubsection{Error Detection}
The ancillary qubits are partitioned into those denoted as $X$ and $Z$ ancilla qubits. These ancilla qubit sets are sufficient for capturing any Pauli error on the data qubits, as $Y$ operators can be treated as a simultaneous $X$ and $Z$ error. The action of the $X$ stabilizer is two-fold: the four neighboring data qubits are forced into a particular state that discretizes any errors that may have occurred on them. Second, the measurement of the $X$ ancilla qubit signals the parity of the number of errors that have occurred on its four neighbors. For example, it yields a $+1$ value if the state of the four neighboring qubits has an even number of $Z$ errors. The same is true of the $Z$ stabilizers -- these track the parity of $X$ errors occurring in the neighboring qubits. If an odd number of errors have occurred in either case, the ancilla qubit measurement will yield a $+1$ value, an event known as a \emph{detection event}~\cite{fowler2012topological}, otherwise these will return values of $0$ or $-1$ depending on convention. We will refer to the ancillary qubits returning $+1$ values as \emph{hot syndromes}. The \emph{error syndrome} of the code is a bit string of length equal to the total number of ancilla qubits, and is composed of all of these measurement values.


\begin{figure*}[t!]
  \centering
  \includegraphics[width=0.9\linewidth]{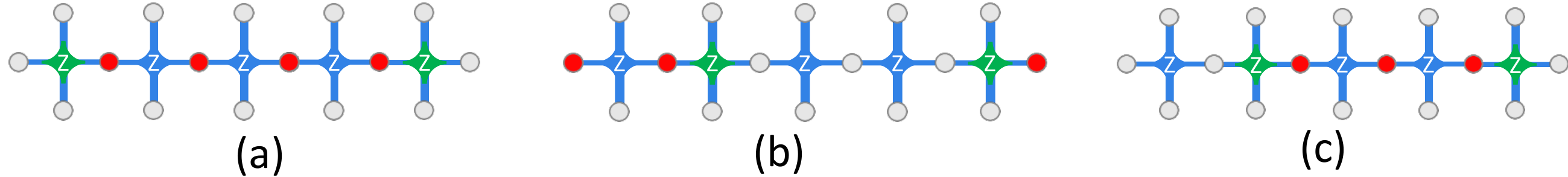}
  \caption{Figure (a) shows a data qubit error pattern spanning across ancillary qubits. Each data qubit experiencing error is indicated in red, and the ancillary qubits returning $+1$ measurement values are indicated in green. Each ancillary qubit that is adjacent to two erroneous data qubits does signal the presence of any errors, as the parity of the data qubit sets are still even. This creates an \emph{error string} that runs from the ancillary qubit on the left of the grid to the one on the right. Decoding must map these $+1$ values to the corresponding set of $4$ data qubit errors that generated it. Figures (b) and (c) show degeneracy in error syndrome generation by surface code data qubit error patterns. The figures depict two distinct sets of data qubit error patterns that both generate the same error syndromes. Both patterns contain the same number of physical data errors, so these patterns are equally likely assuming independence of errors.} 
  \label{fig:error_chain}
\vspace{-0.5cm}
\end{figure*}


Decoding is the process of mapping a particular error syndrome string to a set of corrections to be applied on the device. An example of this process is shown graphically in Figure \ref{fig:surface_mesh}. In this example, the hot syndromes generated by a single data qubit error are marked in red. Each single data qubit error causes the adjacent ancillary qubits 
to return $+1$ values.

A different situation occurs when strings of data qubit errors cross ancillary qubits, as shown in Figure \ref{fig:error_chain}. Here, four consecutive data qubits experience errors which generates hot syndrome measurements on the far left and right of the grid. This is because each ancillary qubit along this chain detects even error parity, so they do not signal the presence of errors. Decoding must be able to pair the two hot syndromes, applying corrections along the chain that connects them.

\subsubsection{Error Detection Can Fail}
Notice that in Figure~\ref{fig:error_chain} (a), if the data qubits on the left and right endpoints of the chain had also experienced errors, none of the ancillary qubits would have detected the chain. This represents a class of undetectable error chains in the code, and specifically occurs when chains cross from one side of the lattice to the other. The result of these chains are physical errors present in the code that cannot be corrected, and are known as \emph{logical errors}, as they have changed the state of the logical qubit. One important characteristic of the surface code is the minimal number of qubits required to form a logical error. This number is referred to as the \emph{code distance, d} of a particular lattice.

\subsection{Quantum Computing Systems Organization}
While qubits are the foundation of a device, a quantum computer must contain many layers of controlling devices in order to interact with qubits. Qubits themselves can be constructed using many different technologies, some of which include superconducting circuits \cite{maslov2,koen5,koen6,barends2014superconducting,Kelly2015}, trapped ions \cite{maslov1,maslov2,maslov3,haffner2005scalable,lekitsch2017blueprint}, and quantum dots \cite{zajac2016scalable}. Controlling these devices is often performed by application of electrical signals at microwave frequencies \cite{chow2012universal,yang2003possible,paraoanu2006microwave,plantenberg2007demonstration}. 


This work focuses on systems built around qubits that require cryogenic cooling to milliKelvin temperatures \cite{hornibrook2015cryogenic}. These systems require the use of dilution refrigerators, and typical architectures involve classical controllers located in various temperature stages of the system. Such a system is described schematically in~\cite{tannu2017taming,hornibrook2015cryogenic}, with a host machine operating at room temperature, and control processors located within the refrigerator itself.

Cryogenically cooled systems present with many design constraints. Specifically, classical controllers located inside the refrigerator are subject to area and power dissipation constraints, as they must not exceed the existing area within a specific temperature stage nor can they dissipate more power than can be cooled by the particular stage \cite{patra2018cryo,sebastiano2017cryo}. Additionally, communication between stages can be costly. Many systems are constructed today using control wiring that scales linearly with the number of qubits in the quantum device, which will prohibit the construction of scalable quantum computers \cite{franke2018rent}. 

\subsection{Classical Control in Quantum Computing Systems}
Inside a cryogenically cooled quantum computer, classical control must be present to perform several different functions. A host processor must handle the expression of high level algorithms and communicate these algorithms to a second controller responsible for the translation of higher level instructions into microcode that represents physical transformations to be performed on the quantum device. The controllers must also appropriately handle the readout of qubit values, and propagate them through the system. The need for fast communication and effective classical processing within the system has motivated much work examining the various constraints and possible engineering solutions to co-locating classical processors throughout these levels \cite{ware2017superconducting}. Feasibility studies have also been conducted \cite{tannu2017cryogenic} as has controller design \cite{tannu2017taming}.


Error correction classical processing requires high bandwidth communication of the measurement values of many qubits on the quantum substrate repeatedly throughout the operation of the device. As a result, not only are instruction streams primarily dominated by quantum error correction operations \cite{levy2009impact,levy2011implications}, but also the classical controller responsible for error correction processing must be tightly coupled to the quantum substrate. If communicating between the quantum substrate and error correcting controller is subject to excessive latencies, the execution of fault tolerant algorithms will be completely prohibited.

\begin{table}[]
\footnotesize
\centering
\begin{tabular}{|l|c|c|c|}
\hline
 & \cellcolor[HTML]{EFEFEF}\# qubits & \cellcolor[HTML]{EFEFEF}\# total gates & \cellcolor[HTML]{EFEFEF}\# T gates \\ \hline
takahashi\_adder & 40 & 740 & 266 \\ \hline
barenco\_half\_dirty\_toffoli & 39 & 1224 & 504 \\ \hline
cnu\_half\_borrowed & 37 & 1156 & 476 \\ \hline
cnx\_log\_depth & 39 & 629 & 259 \\ \hline
cuccaro\_adder & 42 & 821 & 280 \\ \hline
\end{tabular}\vspace{0.1cm}
\caption{Characteristics of the simulated benchmarks.}
\label{table:bench}
\vspace{-0.8cm}
\end{table}

\section{Motivation: Decoding Must be Fast}
\label{sec:fast}
Decoding must be done quickly for the surface code to perform well. During actual computation on a surface code error corrected device, there exist gates called $T$-gates that require knowledge of the current state of errors on the device before they can execute. \footnote{Errors commute and can be post-corrected for other gates, but not $T$-gates.} If decoding is slower than the rate at which syndromes are generated, an algorithm will create a \emph{data backlog}. While the machine is waiting for decoder to process the backlog, more syndrome data is accumulating on the device, which must be processed before executing the subsequent $T$-gate. Over time, this results in latency overhead that is exponentially dependent upon the number of such gates. Specifically, the overhead scales as $(\frac{r_{\text{gen}}}{r_{\text{proc}}})^k = f^k$, where $r_{\text{gen}}$ is the rate of data generation, $r_{\text{proc}}$ is the rate of decoder processing, each in bauds, $f$ is the decoding ratio, and $k$ is the number of $T$ gates in the quantum algorithm. An exponentially slow quantum computer eliminates all of its usefulness. 

\begin{figure}[t!]
  \centering
  \includegraphics[width=0.9\linewidth]{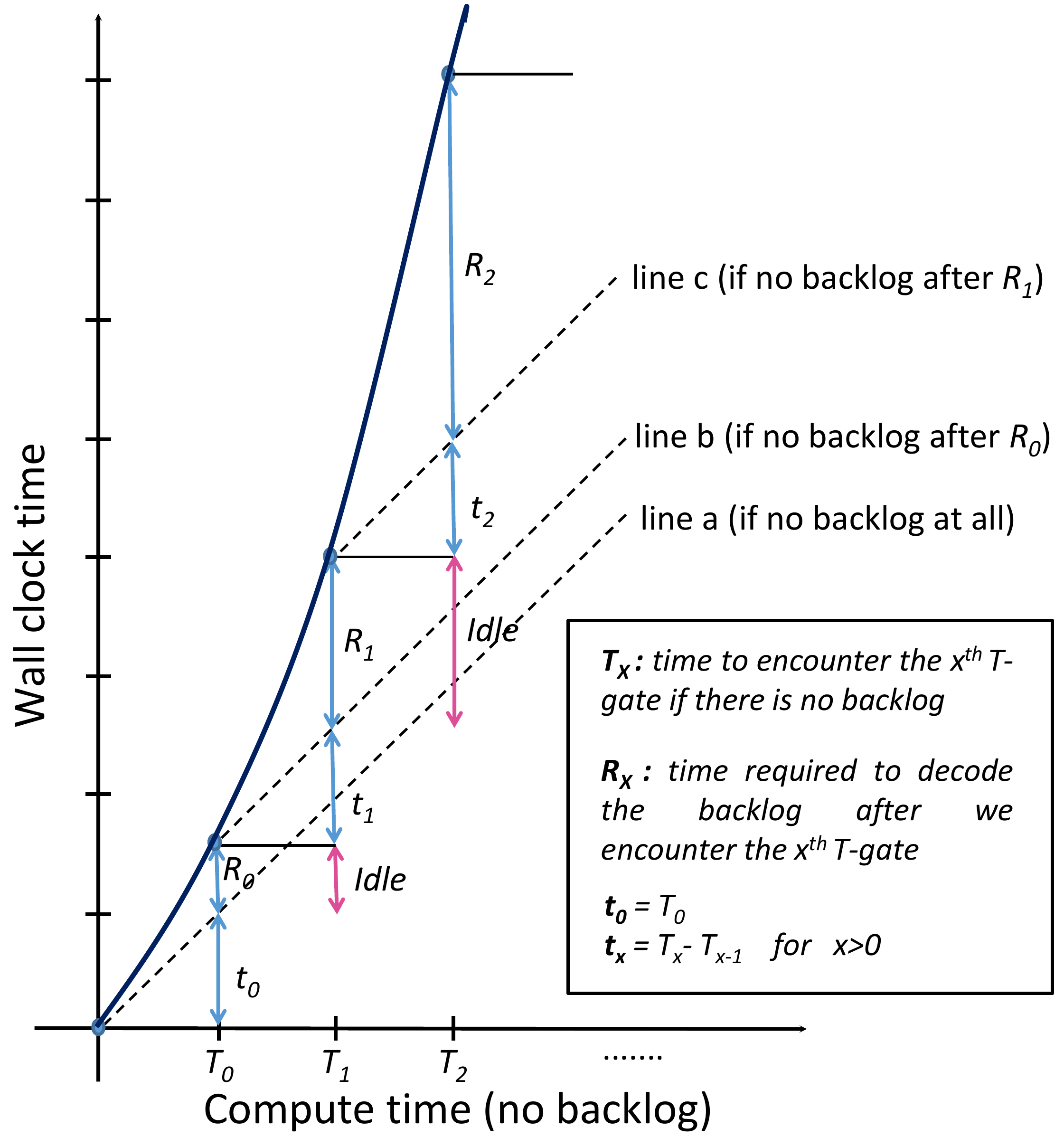}
  \vspace{-0.2cm}
  \caption{Exponential latency overhead when $f=(\frac{r_{\text{gen}}}{r_{\text{proc}}})>1$. X-axis shows the compute time if there is no backlog and y-axis shows the actual wall clock time; if there is no backlog we expect wall clock time to be the same as the compute time (line a). Every time we encounter a T-gate we need to decode all the syndromes up until that gate before we can continue the execution \cite{terhal2015quantum}. When we encounter the first T-gate at time $T_0$, we need to finish the decoding of the data generated during $t_0$ (not all the data is already decoded as decoding rate is slower than data generation rate) and it takes $R_0$ to do that. During $R_0$ where our quantum system is idle, more syndromes are generated and when we encounter the second T-gate at $T_1+R_0$, we need to finish decoding those syndromes in addition to the syndromes generated during $t_1$ before continuing the program execution. {\bf The syndrome data generated during the idle periods is the key reason behind data backlog creation which leads to exponential latency overhead.}}
  \label{fig:backlog}
\vspace{-0.7cm}
\end{figure}

Figure \ref{fig:backlog} shows the exponential latency overhead due to data backlog. The proof of this is summarized as follows (for more details see \cite{terhal2015quantum}): suppose $f > 1$. This implies that there will be a time $t_0$ in the application where we encounter a $T$ gate and must wait for syndrome data to be decoded before continuing. Let $\Delta_{\text{gen}}$ be the amount of time that the machine must stall for processing this data. During this time an additional $D_1 = r_{\text{gen}} \times \Delta_{\text{gen}}$ bits of syndrome data is generated, which can be processed in time $\Delta_{\text{proc}} = r_{\text{gen}}\Delta_{\text{gen}}/r_{\text{proc}} = f\Delta_{\text{gen}}$. The backlog problem begins to be noticeable at this point, where during processing of the first block $D_1$, we generate a \emph{new block} $D_2 = r_{\text{gen}} \times \Delta_{\text{proc}} = fD_1 > D1$ in size. Then, at the next $T$ gate this process repeats, and we again generate a block of data of size $D_3 = fD_2 = f^2D_1$ bits. Hence, by the $k$'th $T$ gate, we generate an overhead of $f^kD_1$ bits to process, exponential in \emph{the decoder's performance ratio}.

As a specific example, consider a multiply-controlled NOT operation on 100 logical qubits from \cite{holmes2018impact}. This algorithm contains $\sim 2356$ gates, of which $686$ are $T$-gates after decomposition. 
Assuming that a syndrome generation cycle time is approximately 400 ns \cite{ghosh2012surface}, and the best prior decoder requires $800$ ns to execute \cite{chamber2018deep}, the ratio $(r_{\text{gen}}/r_{\text{proc}}) = 2$, and the execution time is approximately $10^{196}$ seconds. 

\begin{figure}[t]
  \centering
  \includegraphics[height=5.2cm,width=0.9\linewidth]{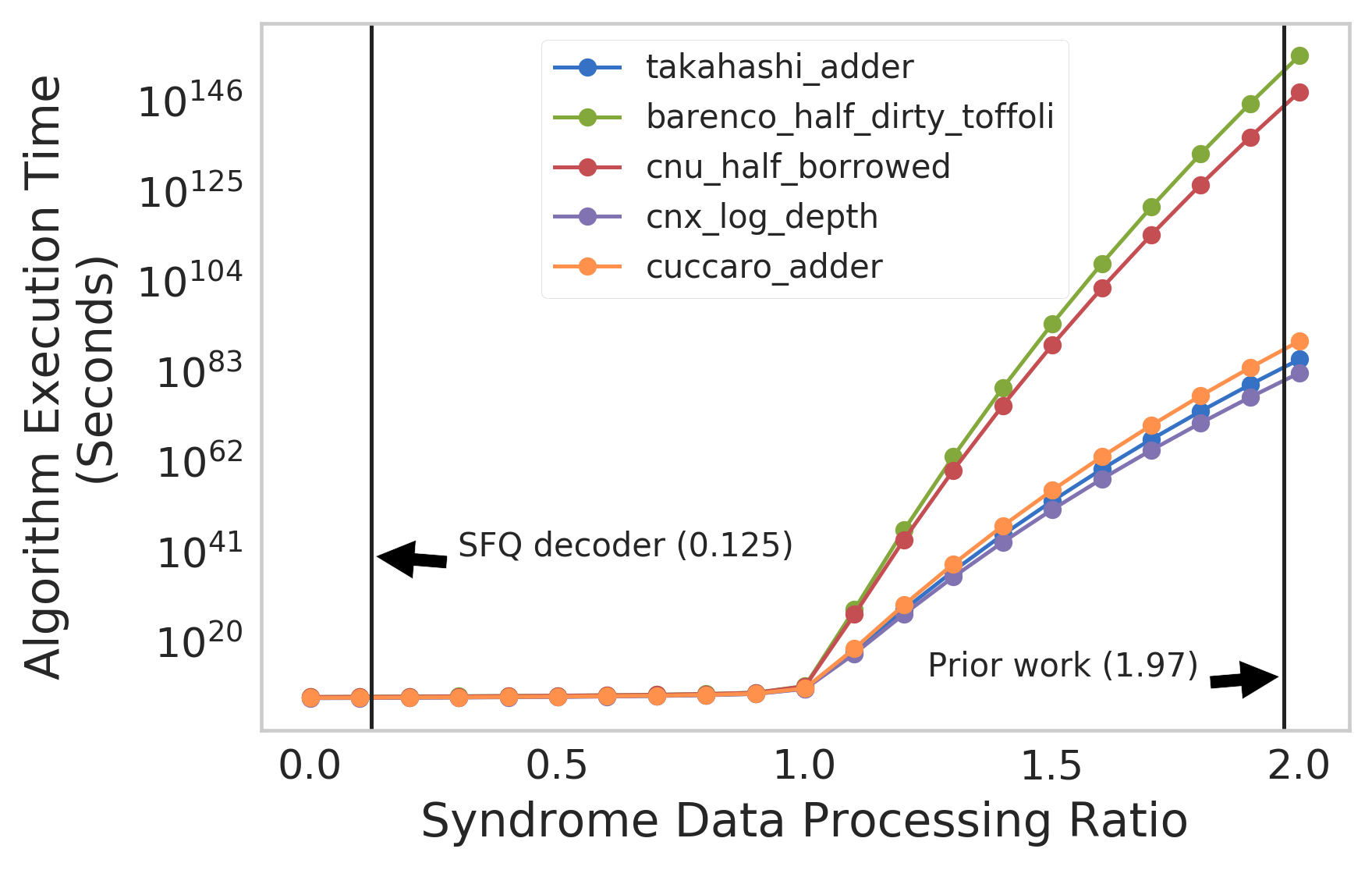}
  \vspace{-0.2cm}
  \caption{Running times of fault tolerant quantum algorithms with decoders of varying efficiency. The X-axis plots $\frac{r_{\text{gen}}}{r_{\text{proc}}}$. 
  To the left of 1, data is processed as fast as it is generated, whereas rates to the right of 1 indicate that the decoder is slower than syndrome data is generated. The $T$-gates require synchronization with the decoder in order to execute. Prior work \cite{chamber2018deep} claims that fast neural network inference decoders can perform inference in $\sim$ 800 ns, which places the decoder at approximately the 1.5 - 2 region for a system generating syndromes in the 400-500ns range. Our decoding results show that time to solution never exceeds 20ns, placing it below 1. Clearly computation becomes intractable quickly for slow decoders.}
  \label{fig:t-overhead-scaling}
\vspace{-0.6cm}
\end{figure}

Figure \ref{fig:t-overhead-scaling} shows a simulation of real quantum subroutines each composed of a different number of $T$ gates as denoted in Table \ref{table:bench}. The exponential overhead scaling shows that as decoders become slower than the rate at which data is being generated (which occurs for ``syndrome data processing ratios" over 1), the overheads quickly become intractable. Regardless of the effectiveness of the decoder, if it operates at a processing ratio higher than 1 then it will impose exponentially high latency overheads on algorithm execution. The algorithms all draw inspiration from \cite{barenco1995elementary}.  Barenco-half-dirty-Toffoli is a logarithmic depth multi-control Toffoli gate using O(n) ancilla bits. It performs the same computation as the ``cnx-log-depth” gate with a different circuit.
The ``cnu-half-borrowed" gives an implementation of a multi-control Toffoli using O(n) dirty ancilla, meaning the initial states of these bits does not need to be known. 
The Cuccaro adder is a linear depth implementation of a reversible A + B adder, i.e. two registers of the specified length added together. It has a carry in and a carry out bit as well. The Takahashi adder is an optimized version of the Cuccaro adder \cite{takahashi2009quantum}.


This is the primary motivation for this work -- the hardware decoder must be able to execute faster than syndrome data are generated as a prerequisite for tractable fault tolerant computation.\label{sec:Motivation}

\section{Related Work}\label{sec:related}

Early work focused on the development of and modifications to the minimum-weight perfect matching algorithm (MWPM) \cite{edmonds1965paths,edmonds1965maximum} to adapt it to surface code decoding \cite{fowler2012towards,fowler2012timing}. This resulted in a claimed constant time algorithm after parallelization \cite{fowler2013minimum}.

Other work has constructed maximum likelihood decoders (MLD) based on tensor network contraction \cite{bravyi2014efficient}. This work is computationally more expensive than minimum-weight perfect matching, but is more accurate.

Neural networks have been explored as possible solutions to the decoding problem as well \cite{koen1,koen2,koen3,koen4,chamber2018deep,varsamopoulos2017decoding,varsamopoulos2018designing,baireuther2019neural,torlai2017neural}. Feed-forward neural networks and recurrent neural networks have been explored in combination with lookup tables to form decoders. The primary distinguishing factor in these systems is that the networks function as \emph{high level decoders} in that they predict both a sequence of error corrections on data qubits along with the existence of a logical error. In this sense, they operate at a higher level than both the MWPM and MLD decoders, seemingly at the cost of execution time with respect to training complexity. 

Lastly, more customized algorithms have been developed specifically targeting the surface code decoding problem, including renormalization group decoders \cite{duclos2010renormalization}, union-find decoding \cite{delfosse2017linear,delfosse2017almost}, and others \cite{wootton2015simple,duclos2010fast}.

\textbf{The primary distinguishing factor of our work is that the decoder design is guided by practical system performance.} Accuracy has been sacrificed in order to achieve quantum advantage. While the proposed decoder design may not achieve logical error suppression at the same order as some other algorithms, the ability to perform the algorithm in SFQ hardware at or exceeding the speed of syndrome generation is achieved, as is satisfaction of system design constraints.

\section{Decoder Overview and Design}\label{sec:decoder}
In this section we describe decoding in terms of a maximum-weight matching problem, followed by details of our approximate decoding algorithm, and demonstrate how we make efficient use of unique features of SFQ gates to implement the algorithm in hardware.

\subsection{Maximum Weight Matching Decoding}
The decoding problem requires that the maximally likely set of error chains be reported as a solution, given a particular error syndrome. This can be formulated as a matching problem. Specifically, given an error syndrome string $S \in \{-1,1\}^n$, we can construct a complete graph on vertices associated with each ancillary qubit that reported an error. The weight of each edge between vertices is proportional to likelihood of a path between these ancillary qubits on the original surface code grid graph. The goal is therefore to find the maximally likely pairing of the syndromes using these weights, one method for doing so is to solve a maximum-weight perfect matching problem.


\subsection{A Greedy Approach}
Our decoding algorithm is based upon a greedy approximation to the maximum-weight matching problem. The algorithm calculates all distances $d(v_i,v_j)$ between vertices and sorts them in ascending order $d_1, d_2, ..., d_{k'}$ where $k' = {k \choose 2}$. All of the corresponding probability weights are calculated, 
transforming this ordering to a descending order of likelihood. Then, for each edge $e$ in descending order, add $e$ to the solution $M$ if it forms a matching. This means that it adds another two distinct vertices into $M$ that were not already present. To account for boundary conditions, we introduce a set of external nodes connected to the appropriate sides of the lattice, and connected to one another with weight $0$. Under this formulation, the algorithm is a 2-approximation of the optimal solution \cite{drake2003simple}. 

\begin{figure}[t!]
  \begin{center}
  \includegraphics[width=0.99\linewidth]{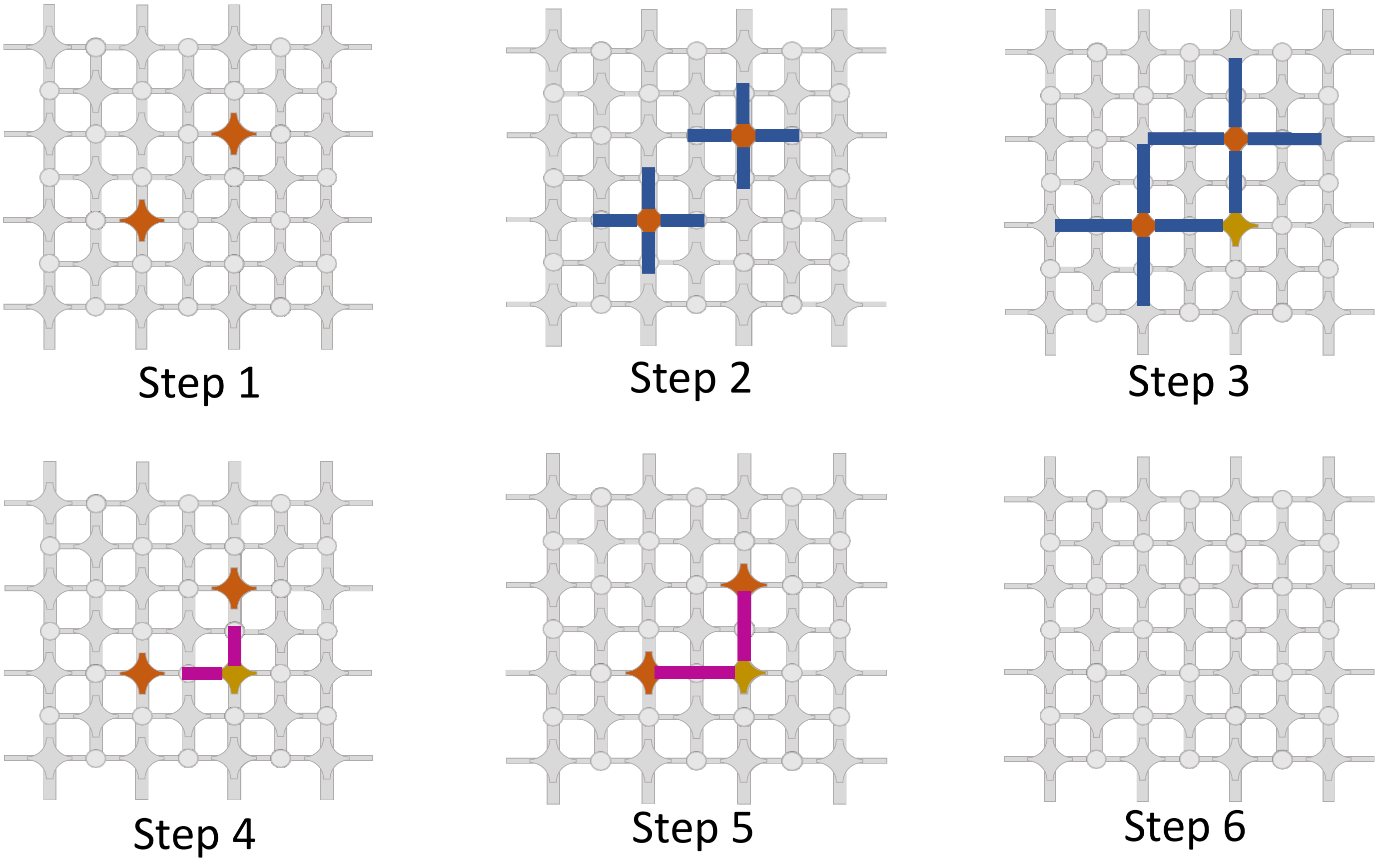}
  \caption{Baseline solution to find the two closest hot syndrome modules. Step1: two decoder modules have ``1'' hot syndrome input. Step2: the hot syndrome modules propagate grow signals. Step3: the grow signals meet at an intermediate module. Step4: the intermediate module sends pair signals in the opposite direction. Step5: pair signals arrive at the hot syndrome modules. Step6: decoding is complete. Note that the decoder modules that receive a pair signal are considered as part of the error chain that has occurred.}
  \vspace{-15pt}
  \label{fig:alg}
  \end{center}
\end{figure}

\subsection{SFQ-Based Decoder}

In this section, we introduce the functional design of our SFQ-based decoder and give some rational for each aspect of its design. As a reminder, Single Flux Quantum is classical logic implemented in superconducting hardware that does not perform any quantum computation. It is a medium used to express our classical algorithm. The decoder is placed above the quantum chip layer; it receives measurement results from ancillary qubits as input, and returns a set of corrections as output. For scalability, our decoder design is built out of a two dimensional array of modules implemented in SFQ logic circuits that we refer to as \emph{decoder modules}. These are connected in a rectilinear mesh topology. Modules are identical and there is one module per each data and ancillary qubit, denoted as \emph{data qubit modules} and \emph{ancilla qubit modules}, respectively. 
Each decoder module has one input called the \emph{hot syndrome input} that comes from the measurement outcome of the physical quantum bits and determines if the module corresponds to a hot syndrome (note that this input can be ``1'' only for ancilla qubit modules). Each module contains one output called the \emph{error output} that determines if the module is contained in the error chain (this output can be ``1'' for all of the decoder modules). In addition, each module has connections to adjacent modules (left, right, up and down).

Our approximate decoder algorithm proceeds as follows. First, the algorithm finds the two modules with ``1'' hot syndrome input, called \emph{hot syndrome modules}, that are closest together.
Next, the algorithm reports the chain of modules connecting them as the correction chain. Finally, it resets the hot syndrome input of the two modules 
 and searches for the next two closest hot syndrome modules. The decoder continues this process until no
module with ``1'' hot syndrome input exists. This is graphically displayed in Fig.~\ref{fig:alg}.

\begin{figure}[t!]
  \begin{center}
  \includegraphics[width=0.99\linewidth]{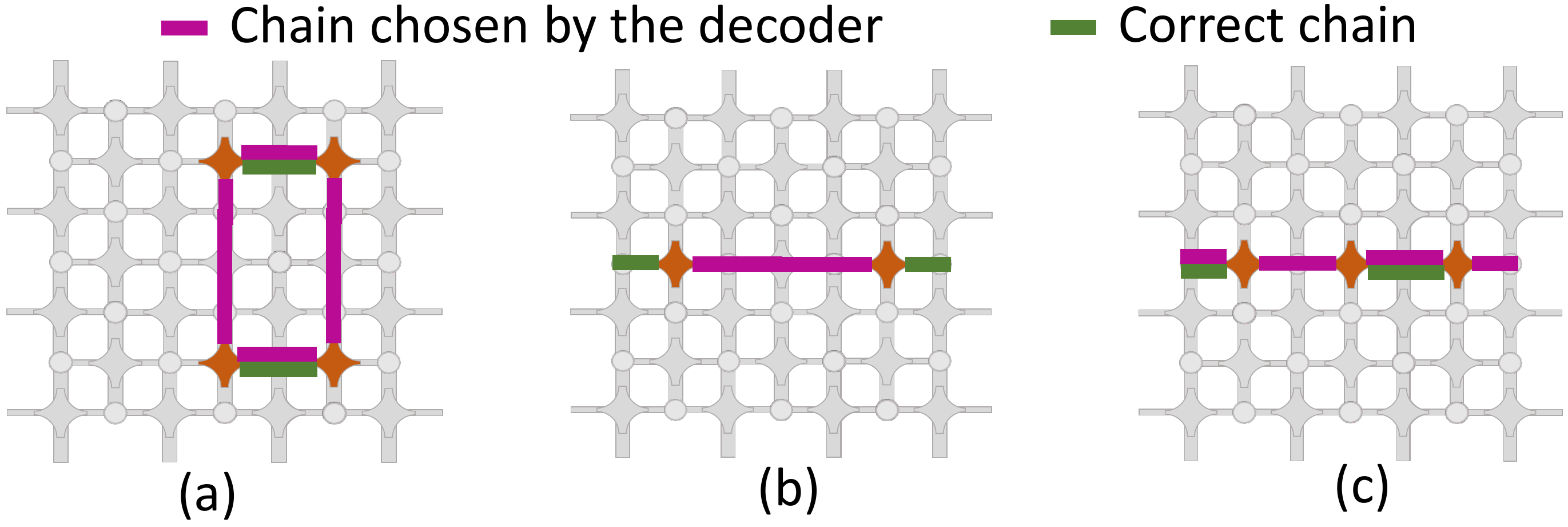}
  \caption{Scenarios where the SFQ decoder chooses the wrong chain where (a) no reset/boundary/equidistant mechanisms are employed, (b) no boundary/equidistant mechanisms are employed, and (c) no equidistant mechanism is employed.}
  \vspace{-25pt}
  \label{fig:increment}
  \end{center}
\end{figure}

{\bf Baseline Solution:} Our baseline design finds the two closest hot syndrome modules as shown in Figure \ref{fig:alg} as follows: 1) every hot syndrome modules sends \emph{grow} signals to all the adjacent modules in all four directions; each adjacent module propagates the grow signal in the same direction. Grow signals propagate one step at each cycle. 2) When two grow signals intersect at an \emph{intermediate module}, we generate a set of \emph{pair} signals and back-propagate these to their hot syndrome origins. 
All of the decoder modules that receive pair signals are part of the error chain. Note that more than one intermediate module might exist, however, only one of them is effective and sends the pair signals. For example in Figure \ref{fig:alg}, two intermediate modules receive the grow signals, and the decoder is hardwired to be effective (ineffective) when it receives grow signals from up and left directions (down and right directions). Intermediate module refers to the effective one. The baseline solution does not show accuracy or pseudo-threshold behavior and demonstrates poor logical error rate suppression, see the incremental results presented in Section \ref{sec:evaluation} in Figure~\ref{fig:results}. 

{\bf Reset Mechanism:} One flaw of the baseline system is the lack of a mechanism to reset the decoder modules after two hot syndrome modules are paired. Grow signals of the paired modules continue to propagate, potentially causing these modules to pair incorrectly with other hot syndrome modules, ultimately resulting in an incorrect error chain reported. Figure \ref{fig:increment} (a) shows an incorrect matching due to this behavior. To mitigate this, we add a reset mechanism that resets the decoder modules each time hot syndrome modules are paired and the error chain connecting them is determined. Adding the reset mechanism to the baseline system improves the performance somewhat, but does not yet achieve tolerable accuracy. 

{\bf Boundary Mechanism:} Another explanation for the low performance of the baseline solution is that it never pairs hot syndrome modules with boundaries. For example, if two hot syndrome modules are far from each other but are close to boundaries, the error chain with the maximum likelihood is the one that connects the hot syndrome modules to the boundaries. Figure \ref{fig:increment} (b) shows this behavior occurring on a machine.
We implement a mechanism that enables pairing the hot syndrome modules with boundaries. To do this, we add decoder modules that surround the surface boundaries called \emph{boundary module} (one per each quantum bit located at a boundary). Our solution treats boundary modules as hot syndrome modules but they do not grow and can pair only with non-boundary modules. Note that when two modules are paired, the hot syndrome input of only the non-boundary modules is reset; boundary modules are always treated as hot syndrome modules. Adding the boundary mechanism to the baseline solution augmented with the reset mechanism further increases the accuracy of the decoder. 
 
 {\bf Equidistant Mechanism:} Finally, the last major reason for inefficiency of the baseline is that it does not properly handle the scenarios in which multiple hot syndrome modules are spaced within equal distances of one another, resulting in a set of pairs that are all equally likely. The baseline solution augmented with reset and boundary mechanisms works properly only if no non-boundary hot syndrome module has an equal distance to more than one other hot syndrome module; otherwise the solution pairs it with all the hot syndrome modules with equal distance. However, this is not the desired output. We need a more intelligent solution to break the tie in the aforementioned scenario, and pair the hot syndrome module to only one other module. This is shown in Figure \ref{fig:increment} (c).
 
 To resolve these equidistant degenerate solution sets, we introduce a request -- grant policy that allows for the hardware to choose specific subsets of these pairs to proceed. 1) Similar to the baseline solution, the non-boundary hot syndromes first propagate grow signals. 2) An intermediate module receives two grow signals from two different directions, and it sends \emph{pair\_request} signals in the opposite directions. Pair\_request signals continue to propagate until they arrive at a module with ``1'' hot syndrome input. 3) The modules with ``1'' hot syndrome input send \emph{pair\_grant} signals in the opposite direction of the received pair\_request signals. Note that multiple pair\_request signals might arrive at a module with ``1'' hot syndrome at the same time, but it gives grant to only one of them. 4) An intermediate module receives pair\_grant signals from two different directions and sends pair signals in the opposite directions. 5) Pair signals continue to propagate until they arrive at a module with ``1'' hot syndrome input. Boundary modules do not send grow signals but they send pair\_request signals when they receive grow signals; they also send pair signals when they receive pair\_grant signals.


\section{Implementation}
\label{sec:implementation}

\subsection{SFQ Implementation of Greedy Decoding}

SFQ is a magnetic pulse-based fabric with switching delay of \emph{1ps} and energy consumption of $10^{-19}$J per switching. In addition, availability of superconducting microstrip transmission lines in this technology makes it possible to transmit picosecond waves with half of speed of light and without dispersion or attenuation. The combination of these properties together with fast two-terminal Josephson junctions, makes this technology suitable for high speed processing of digital information \cite{volkmann2013experimental,kirichenko2011zero,herr2011ultra,takeuchi2013adiabatic,likharev1991rsfq}. SFQ logic families are divided into two groups: ac-biased and dc-biased; Reciprocal Quantum Logic (RQL) \cite{herr2011ultra}, and Adiabatic Quantum Flux Parametron (AQFP) \cite{takeuchi2013adiabatic} are in the first group, and Rapid Single Flux Quantum (RSFQ) \cite{likharev1991rsfq}, Energy-efficient RSFQ (ERSFQ) \cite{kirichenko2011zero}, and energy-efficient SFQ (eSFQ) \cite{volkmann2013experimental} are examples of the second group. The dc-biased logic family with higher operation speed (as high as 770GHz for a T-Flip Flop (TFF) \cite{chen1999rapid}) and less bias supply issues are more popular than ac-biased logic family. 

Our algorithm requires modules to propagate signals one step at each cycle. One approach to implement our algorithm is to use synchronous elements such as flip-flops in decoder modules. However, standard CMOS style flip-flops are very expensive in SFQ logic (e.g., one D-Flip-Flop occupies $72.4 \times$ more area and consumes $117 \times$ more power compared to a 2-input AND gate). On the other hand, SFQ gates have a unique feature that we utilize to implement our algorithm without flip-flops. Unlike CMOS gates, most of the SFQ gates (expect for mergers, splitters, TFFs, and I/Os) require a clock signal to operate \cite{pasandi2018sfqmap}. Thus, we do not need to have flip-flops and signals can propagate one SFQ gate at each cycle.

As described earlier, our decoder requires resetting the decoder modules each time two hot syndrome modules are paired. We have a global wire that passes through all the modules and is connected to each module using splitter gates. Thus, if we set the value of the global wire, all of the decoder modules receive the reset signal at the same time, as the splitter gates do not require clock signals to operate. If a module receives a reset signal, it blocks the module inputs using 2-input AND gates (one input is \emph{module\_input} and the other input is $\overline{Reset}$). In order to reset a decoder module completely, we need to block the module inputs for as many cycles as the depth of our SFQ-based decoder because the SFQ gates work with clock cycles and one level of gates is reset at each cycle. Thus, we use a simple circuit to keep the reset signal ``1'' for as many cycles as the circuit depth. In each module, we pass the reset signal that comes from the global wire to a set of $m$ cascaded buffer gates where $m$ is the circuit depth, and the module inputs are blocked if the reset signal that comes from the global wire is ``1'' or at least one of the buffers has ``1'' output.

\begin{figure}[t!]
  \begin{center}
  \includegraphics[width=0.99\linewidth]{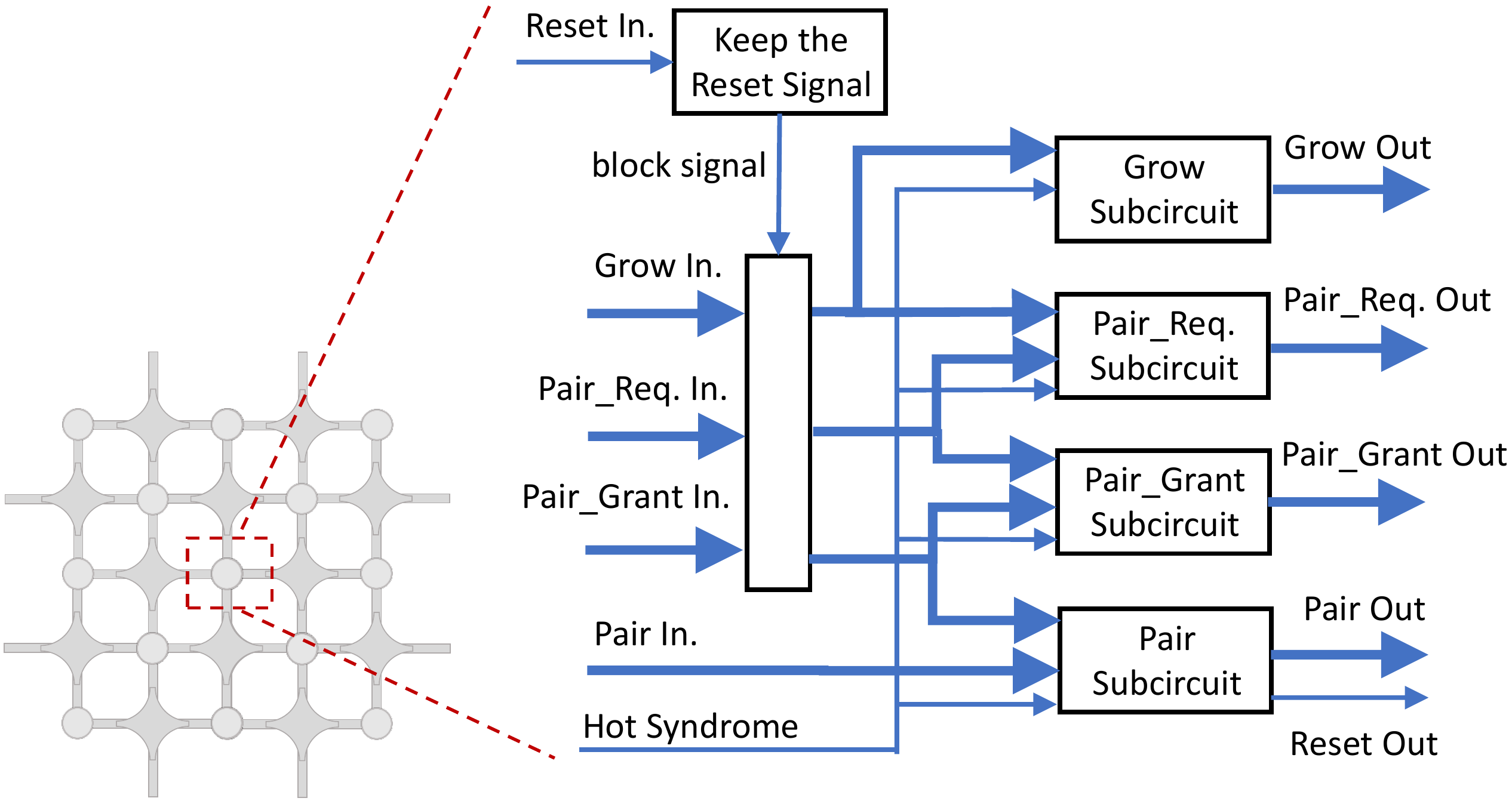}
  \caption{Overview of decoder module microarchitecture.}
  \vspace{-20pt}
  \label{fig:overview}
  \end{center}
\end{figure}

\subsection{Datapath and Subcircuit Design}
Figure \ref{fig:overview} shows an overview of our decoder module microarchitecture. Our decoder consists of five main subcircuits.

{\bf Grow Subcircuit:} this subcircuit receives hot syndrome input and 4 grow inputs (from 4 different directions), and produces 4 grow output signals. Grow outputs are ``1'' if the hot syndrome input is ``1'' or if the module is passing a grow signal generated by another module. 

{\bf Pair\_Req Subcircuit:} this subcircuit is responsible for setting the value of pair\_request outputs which are ``1'' if two grow signals meet at an intermediate module or if the module is passing a pair\_request signal that arrived at one of its input ports. The module does not pass the pair\_request input signal if the hot syndrome input is ``1''; in that case, the module generates a pair\_grant signal instead. 

{\bf Pair\_Grant Subcircuit:} this module determines the value of pair\_grant outputs which are ``1'' if the module is a hot syndrome module and gives grant to a pair\_request signal, or if the module is passing a pair\_grant input signal to the adjacent module. 

{\bf Pair Subcircuit:} this subcircuit sets the value of pair outputs which are ``1'' if two pair\_grant signals meet at an intermediate module or if a pair input signal is ``1'' and the hot syndrome input is not ``1''. If both the pair input and hot syndrome input are ``1'', the module does not pass the pair signal and instead generates a global reset signal that reset all of the decoder modules and also resets the hot syndrome input. Note that the reset signal resets everything expect the subcircuit responsible for passing the pair signals because it is possible that the intermediate module does not have equal distance from the paired hot syndrome modules and we do not want to stop the propagation of all the pair signals in the system when the closer module receives a pair signal (while the farther module has not received a pair signal yet). 

{\bf Reset Subcircuit:} this subcircuit is responsible to keep the reset signal ``1'' for as many cycles as the depth of our circuits. The depth is 5 in our circuits, thus reset subcircuit blocks grow, pair\_req and pair\_grant inputs for 5 cycles in order to reset the module.

\section{Methodology}\label{sec:methodology}
{\bf Simulation Techniques:}
In order to effectively benchmark the performance of a stabilizer quantum error correcting code, techniques must be used to simulate the action of the code over many cycles. This is referred to elsewhere in literature as \emph{lifetime simulation}~\cite{varsamopoulos2018designing}, or simply Monte Carlo benchmarking. We constructed a simulation environment that simulates the action of the stabilizer circuits shown in Figure~\ref{fig:surface_stabilizers}. A cycle refers to one full iteration of the stabilizer circuit. At each step within the cycle, errors are stochastically injected into the qubits and propagated through the circuits. Ancillary qubits are measured, and the outcomes are reported in the error syndrome. This syndrome is then communicated directly to the decoder simulator, which returns the corresponding correction. The correction is applied and the surface is checked for a logical error. The ratio of the number of logical errors to the number of cycles run in simulation is used as the primary performance metric.

{\bf Evaluation Performance Metrics:}
\label{sec:metrics}
In our evaluations, we use the stabilizer circuits shown in Figure~\ref{fig:surface_stabilizers} as the primary benchmark. These circuits are replicated for every ancillary qubit present in a surface code lattice. Many different lattices are also analyzed, ranging in size from code distances 3 to 9.

As performance metrics, we focus on \emph{accuracy thresholds} and \emph{pseudo-thresholds}. The former is the physical error rate at which the code begins to suppress errors effectively across multiple code distances. Below this threshold, the logical error rate $P_L$ decreases as the code distance $d$ increases. Above threshold these relationships invert, and $P_L$ grows with $d$ due to decoder performance: the presence of many errors causes the decoding problem to become too complex. In many cases, this leads to corrections that complete what would have otherwise been short error chains, forming logical errors, a process that amplifies as code distances increase.

Pseudo-threshold refers to the performance of a single code distance, and is the physical error rate at which the logical error rate is equal to the physical rate, i.e. $P_L = p$. This can be (and often is) different across different code distances. Better error correcting codes will have higher pseudo-threshold values, as well as higher accuracy thresholds. 

\begin{figure*}[t]
    \centering
    \begin{tabular}[b]{c}
    \includegraphics[width=.27\linewidth]{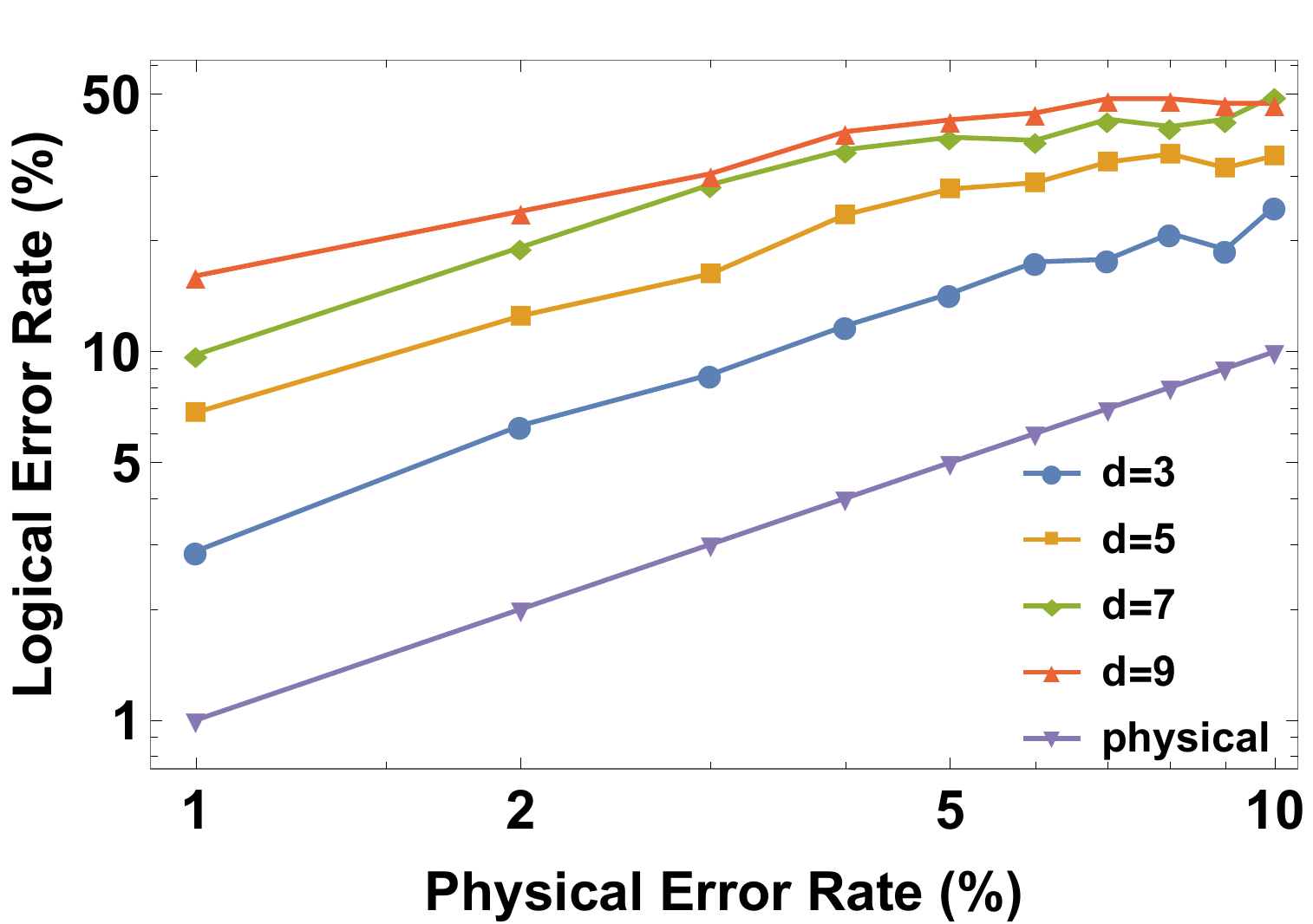} 
        \label{fig:base_nb_nr}\\
    \small Baseline design
  \end{tabular} \qquad
  \begin{tabular}[b]{c}
    \includegraphics[width=.27\linewidth]{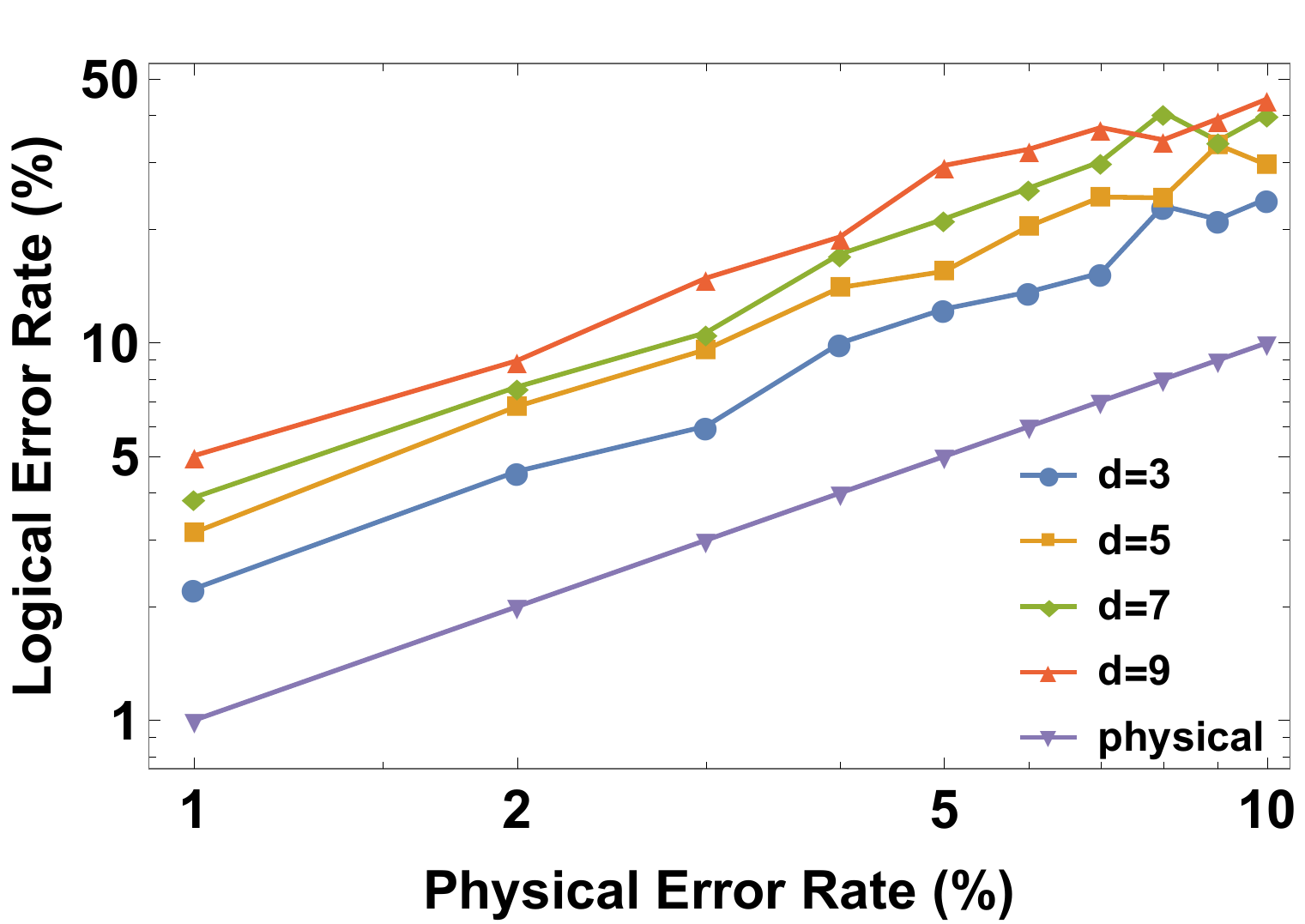} 
        \label{fig:baseline_nb}\\
    \small Adding resets
  \end{tabular} \qquad
  \begin{tabular}[b]{c}
    \includegraphics[width=.27\linewidth]{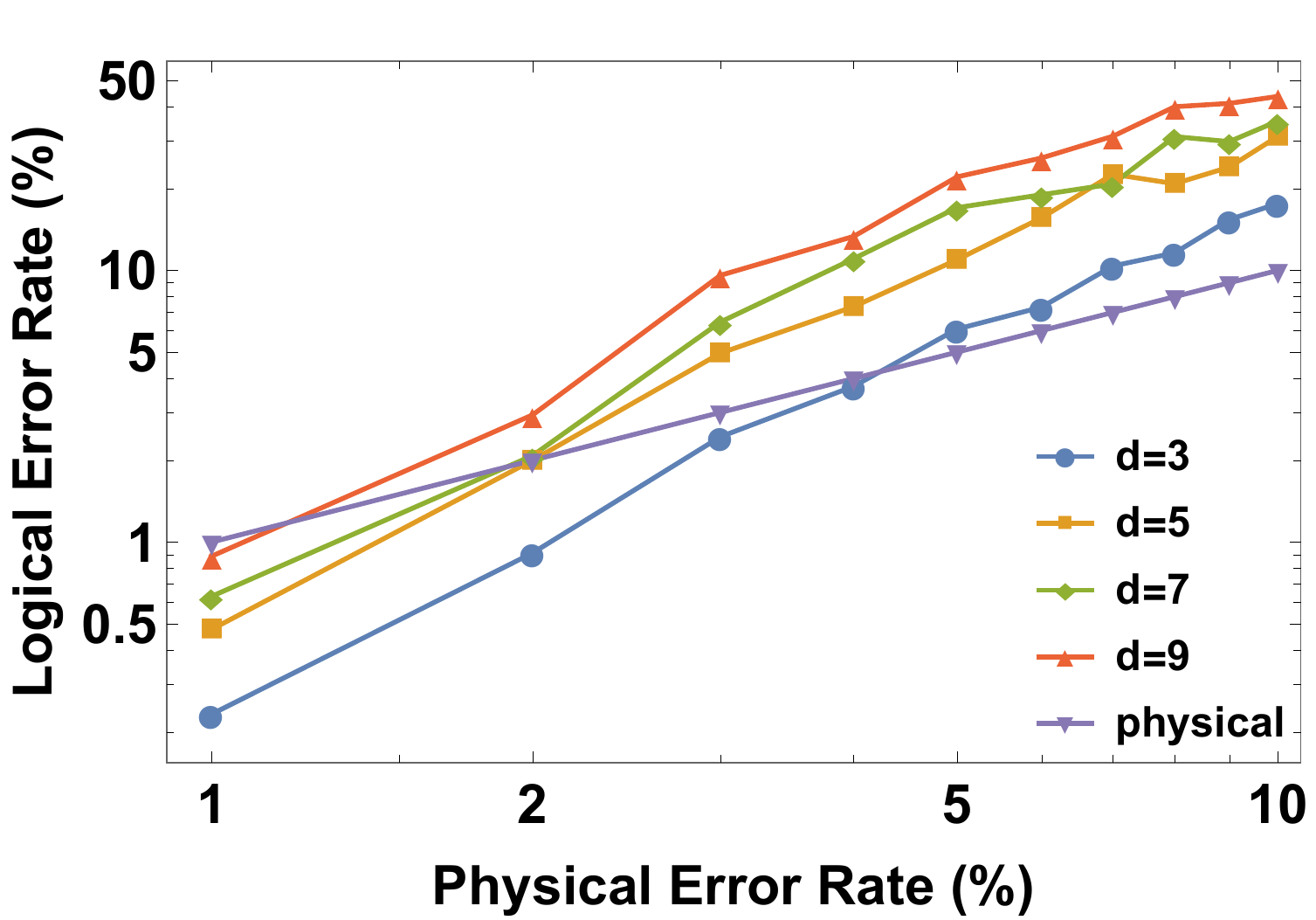} 
        \label{fig:full_base}\\
    \small Adding resets and boundaries
  \end{tabular}
    \label{fig:incremental}
    \vspace{-10pt}
\end{figure*}

\begin{figure*}[t!]
  \centering
  \begin{tabular}[b]{c}
    \includegraphics[width=.27\linewidth]{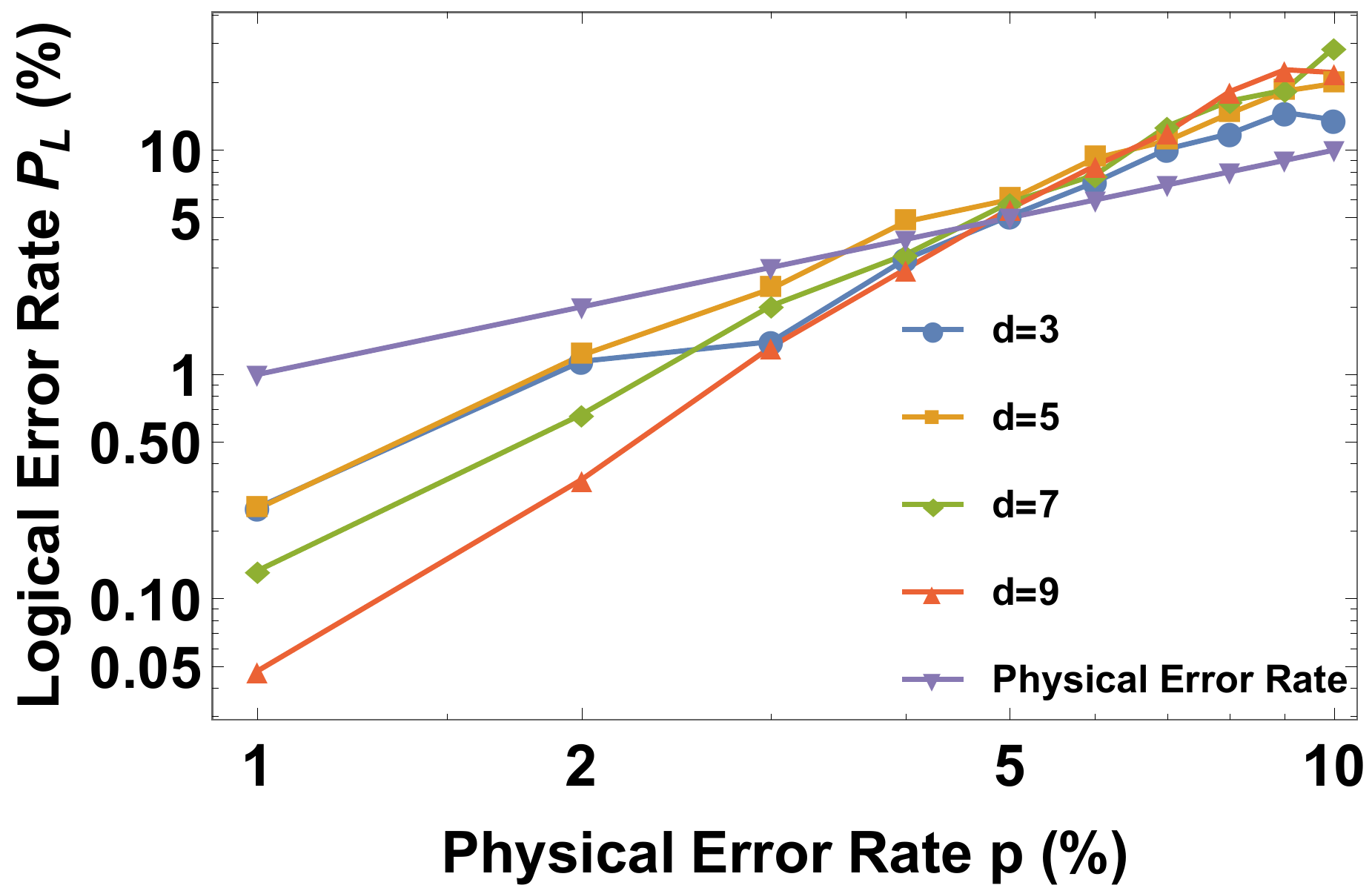} 
        \label{fig:accuracy_threshold}\\
    \small (a) Final design
  \end{tabular} \qquad
  \begin{tabular}[b]{c}
    \includegraphics[width=.27\linewidth]{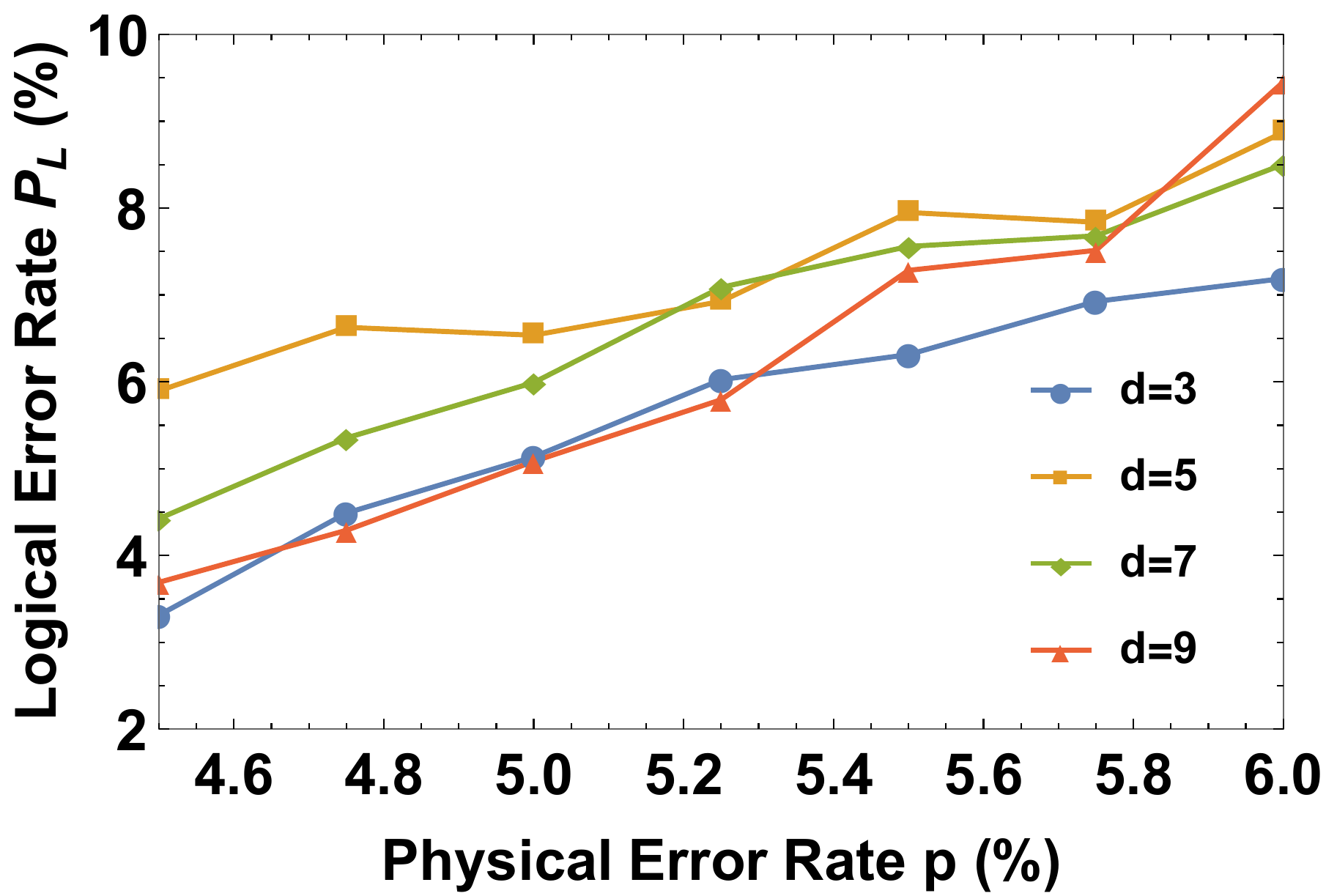} 
        \label{fig:near_threshold}\\
    \small (b) Zoomed in final design
  \end{tabular} \qquad
  \begin{tabular}[b]{c}
    \includegraphics[width=.27\linewidth]{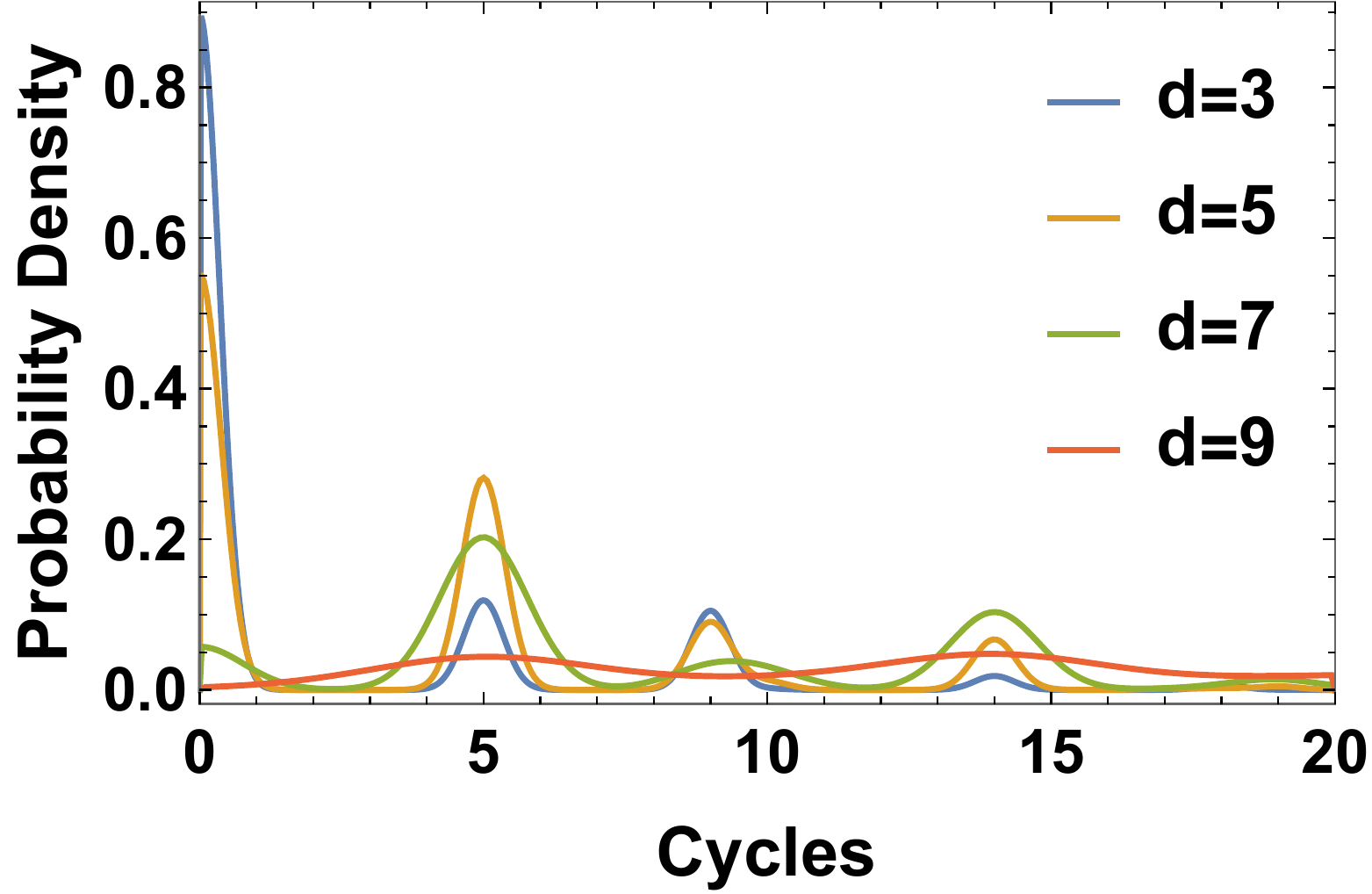} 
        \label{fig:cycle_kernels}\\
    \small (c) Final design probability densities
  \end{tabular}
  \caption{Top row: Logical error rate performance of each incremental design step. The addition of resets and boundaries each contribute heavily to the realization of pseudo-thresholds, and have a dramatic effect on reducing the minimum achievable logical error rates for each code distance. Bottom row: Results for our final design, including support for reset, boundary, and equidistant mechanisms. (a) Error rate scaling for the proposed decoder. An accuracy threshold is evident at approximately $5\%$ physical error rate, while pseudo-thresholds span the range from $\sim3.5\%$ -- $5\%$. (b) Logical error rates near the $5\%$ physical error rate value.(c) Truncated unnormalized estimated probability distributions for the execution cycles required by each code distance in simulation. Window shows up to 20 cycles for comparison across code distances. Notice that while distances $3, 5, 7$ display peaks centered at $0, 5, 9,$ and $14$ cycles.}
  \label{fig:results}
  \vspace{-5pt}
\end{figure*}

{\bf Error Models:}
The Monte Carlo simulation environment requires a model of the errors on the quantum system. We choose to focus on the \emph{depolarizing channel} model \cite{MikenIke,fowler2012surface,lidar2013quantum,terhal2015quantum}, parameterized by a single value $p$: Pauli $X, Y,$ and $Z$ errors occur on qubits with probability $p/3$. During simulation Pauli errors are sampled i.i.d for injection on each data qubit. 
We present analysis of a variation of the model, the \emph{pure dephasing channel} \cite{delfosse2017almost,delfosse2017linear} comprised solely of Pauli $Z$ errors occurring on qubits with probability $p$. The decoder will be operated symmetrically for both $X$ and $Z$ errors, allowing for simple extrapolation from these results.

{\bf Single Flux Quantum Circuit Synthesis:}
\label{sec:sfq_cells}
\begin{table}[t]
\footnotesize
\centering
\begin{tabular}{@{\extracolsep{4pt}}|l|c|c|c|@{}}
\hline\hline
Cell & Area ($\mu$m$^2$) & JJ Count & Delay (ps)\\ \hline
AND2 & $4200$ & $17$ & $9.2$\\
OR2 & $4200$ & $12$ & $7.2$\\
XOR2 & $4200$ & $12$ & $5.7$\\
NOT & $4200$ & $13$ & $9.2$\\
DRO DFF & $3360$ & $10$ & $5.0$\\
\hline\hline
\end{tabular}\vspace{0.1cm}
\caption{The library of ERSFQ cells and corresponding characteristics used for synthesizing the circuit into SFQ hardware. Josephson Junction count is listed in the second column.}
\label{table:cells}
\vspace{-30pt}
\end{table}
An ERSFQ library of cells is used in this paper to reduce the total power consumption (including the static and dynamic) of the surface code decoder as much as possible. Table \ref{table:cells} lists characteristics of this library. As seen, this library contains four logic gates including AND2, OR2, XOR2, and NOT, and it has a Destructive Read-Out D-Flip-Flop (DRO DFF) cell. Area of all logic cells are the same and it is equal to 4200 $\mu$m$^2$. However, area of the DRO DFF is less than the area of these gates (3360 $\mu$m$^2$). DRO DFFs are different from standard CMOS style flip-flips: they are specially designed for SFQ circuits and are usually used for path balancing. In Table~\ref{table:cells}, the total number of Josephson junctions (as a measure of complexity and cost) used in designing each gate together with the intrinsic delay of each cell is reported.

The decoder circuit and its sub-circuits are synthesized by employing ERSFQ specific logic synthesis algorithms and tools \cite{pasandi2018sfqmap,pasandi2019pbmap,pasandi2019balanced}. These algorithms are designed to reduce the complexity of the final synthesized and mapped circuits in terms of total area and Josephson junction count ($\#$JJs). This is achieved by reducing the required \emph{path balancing} DFF count for realizing these circuits. This means that in a Directed Acyclic Graph (DAG) that represents an SFQ circuit, length of any path from any primary input to any primary output in terms of the gate count should be the same. Please note that for correct operation of dc-biased SFQ (including ERSFQ) circuits, these circuits should be fully path balanced. In most of the SFQ circuits this property does not hold in the beginning. Therefore, some path balancing DFFs should be inserted into shorter paths to maintain the full path balancing property. In the algorithms we employed for mapping these circuits, a dynamic programming approach is used to ensure minimization of the total number of DFFs to maintain the balancing property. In addition, a depth minimization algorithm together with path balancing is employed \cite{pasandi2018sfqmap} to reduce the logical depth (length of the longest path from any primary input to any primary output in terms of the gate count) of the final mapped circuit. This helps to reduce the latency of the mapped SFQ circuit. 
As mentioned before, SFQ logic gates are pulsed-based, meaning that the presence of a pulse represents a logic-``1" and the absence of a pulse represents a logic-``0". Each gate is clocked, and as an example the SFQ NOT gate behaves as follows. After the clock pulse arrives, when there are no input pulses, a pulse is generated at the output of the gate representing a ``1". On the other hand, when there is an input pulse, no pulses are generated at the output, meaning a ``0". Each pulse is a single quantum of magnetic flux $(\phi_0=\frac{h}{2e}=2.07$mV$\times$ps) \cite{likharev1991rsfq}. To simulate the SFQ circuits for verifying their correct functionality, we use the Josephson simulator (JSIM) \cite{delport2019josim}.


\section{Evaluations}\label{sec:evaluation}

In this section we evaluate the performance of our proposed decoder design, both in terms of circuit characteristics including power, area, and latency, as well as error correction performance metrics of accuracy and pseudo-thresholds. 
We also analyze the execution time of our system, relying upon described operating assumptions and circuit synthesis results. 

\begin{table*}[t!]
\footnotesize
\centering
\begin{tabular}{@{\extracolsep{4pt}}|l|c|c|c|c|@{}}
\hline\hline
Circuit & Logical Depth & Latency (ps) & Total Area ($\mu$m$^2$) & Power Consumption ($\mu$W)\\  \hline
AND\_GATE & $1$ & $9.20$ & $4200$ & $0.026$\\
OR\_GATE & $1$ & $7.20$ & $4200$ & $0.026$\\
OR\_GATE\_7\_INPUTS & $3$ & $21.60$ & $38640$ & $0.338$\\
NOT\_GATE & $1$ & $9.20$ & $4200$ & $0.026$\\
Pair\_Grant Subcircuit & $5$ & $85.60$ & $338520$ & $3.38$\\
Pair Subcircuit & $5$ & $96.00$ & $347760$ & $3.51$\\
Pair\_Req./Grow Subcircuit & $5$ & $96.00$ & $447720$ & $4.55$\\
Full\_Circuit & $6$ & $162.72$ & $1279320$ & $13.08$\\
\hline\hline
\end{tabular}\vspace{0.1cm}
\caption{Experimental synthesis results for the SFQ Decoder. Shown are all gates utilized in the synthesis, as well as submodules that comprise the main circuit. Pair\_Req. and Grow subcircuits have been combined into a single subcircuit.}
\label{table:synth_results}
\vspace{-25pt}
\end{table*}

\begin{table}[t!]
\footnotesize
\centering
\begin{tabular}{@{\extracolsep{4pt}}|c|c|c|c|@{}}
\hline\hline
Code Distance & Max & Average & Standard Deviation\\  \hline
$3$ & $3.74$ & $0.28$ & $0.58$\\
$5$ & $9.28$ & $0.72$ & $1.09$\\
$7$ & $14.2$ & $2.00$ & $1.99$\\
$9$ & $19.2$ & $3.81$ & $3.11$\\
\hline\hline
\end{tabular}\vspace{0.1cm}
\caption{Decoder execution time in nanoseconds across each code distance studied and across all simulated error rates.} 
\label{table:cycle_stats}
\vspace{-30pt}
\end{table}

{\bf Threshold Evaluations:}
To gauge the performance of our design, we use the threshold metrics described in Section \ref{sec:metrics}. Figure~\ref{fig:results} (a) shows the central performance result, while the top row of Figure~\ref{fig:results} shows the effect of all of the incremental design decisions on the overall performance. This evaluation simulates the performance of the decoder across a range of physical error rates. 
A pseudo-threshold range of between $3.5\%$ and $5\%$ is observed, and an accuracy threshold appears at approximately the $5\%$ error rate. For code distance $5$, the pseudo-threshold is below the accuracy threshold. This highlights the difference between these metrics -- an error correcting protocol like the surface code can perform well even though particular code distances may still be amplifying the physical error rates (i.e. $P_L > p$). It is important to consider both types of thresholds when evaluating decoder performance.

An interesting behavior is observed for code distance $d=3$. This lattice performs at or surpassing the performance of all other lattices from the $3\%$ physical error rate and above. Below this point, the lattice begins to taper off, and ultimately it converges with the distance 5 lattice. 
Boundary conditions were highly prioritized in our design, causing this effect. In particular, the decoder is designed such that error chains that terminate at the boundaries are more likely to be correctly identified than other patterns. This choice was made as smaller lattices are more dominated by these edge effects than larger lattices. The smallest lattice in our simulations shows this anomalous behavior, as it contains a disproportionate amount of boundary patterns. In larger lattices, syndromes are less likely to terminate in boundaries, reducing this effect.


Figure~\ref{fig:results} (b) highlights the desired threshold behavior. Examining the $6\%$ error rate, code distance $9$ is outperformed by code distance $7$. Moving to the lowest physical error rate in the window, we find that the lattices perform in the order $d=9$, $d=3$, $d=7$, and $d=5$, ordered from lowest to highest logical error rate. Barring the anomalous $d=3$ behavior described above, this is exactly accuracy threshold indicative of successful error correction performance.

\begin{table}
\footnotesize
\centering
\begin{tabular}{@{\extracolsep{4pt}}|c|c|c|c|c|@{}}
\hline\hline
Code Distance & 3 & 5 & 7 & 9 \\  \hline
$c_2$ & $0.650$ & $0.429$ & $0.306$ & $0.323$\\
\hline\hline
\end{tabular}\vspace{0.1cm}
\caption{Empirical parameter estimation given a model of the form $P_L \approx c_1 (p/p_{\text{th}})^{c_2 \cdot d}$. Shown are estimated $c_2$ parameter values.}
\label{table:scaling}
\vspace{-30pt}
\end{table}

{\bf Performance Analysis:}
\label{sec:scaling}
To quantify the approximation factor of our design we compare the performance to that of an ideal decoder by fitting to an exponential analytical model. The achievable error rates by the surface code ideally can be described by $P_L \approx 0.03(p/p_{\text{th}})^d$ \cite{fowler2012surface} when a minimum weight matching decoder is used in software. Using a model of the form $P_L \approx c_1 (p/p_{\text{th}})^{c_2 \cdot d}$, we fit values of $c_1, c_2$ for each code distance at physical error rates below accuracy threshold, and collect $c_2$ values in Table~\ref{table:scaling}. $C_2$ coefficients describe the \emph{effective code distance} for our system, and capture the approximation factor we introduce. For code distances 3 and 5, we find that the approximate decoder is roughly $65\%$ and $43\%$ of the optimal distance respectively. This is the trade-off made by our system in order to fit the timing and physical footprint requirements of the system. 

\begin{figure}[t!]
  \begin{center}
  \includegraphics[width=0.95\linewidth, clip]{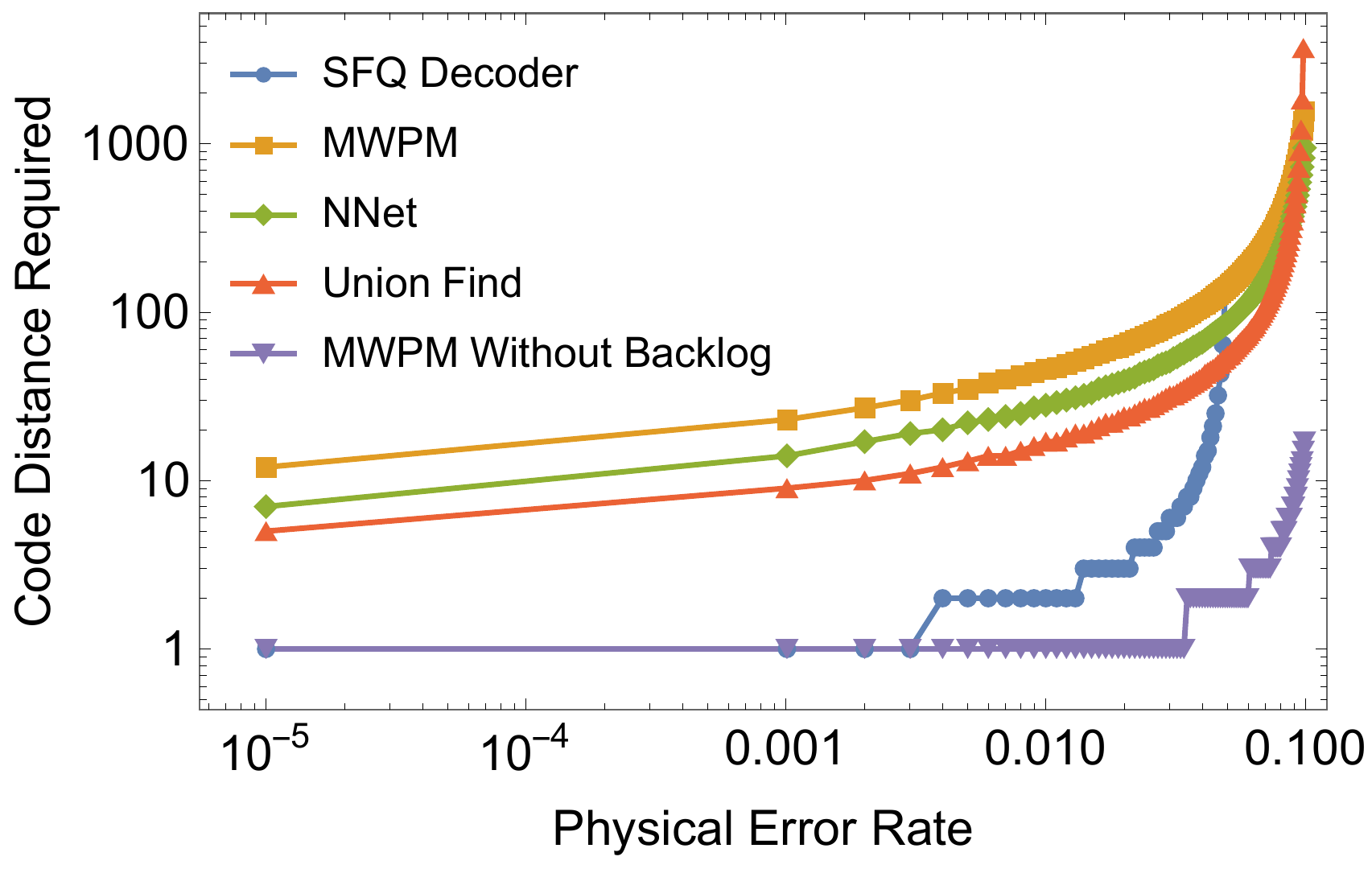}
  \vspace{-5pt}
  \caption{Comparison of required code distances of different decoders to execute an algorithm consisting of 100 T-gates. Compared are the SFQ Decoder, minimum weight perfect matching decoder (MWPM) \cite{fowler2012surface}, neural network decoder \cite{baireuther2019neural},union find decoder \cite{delfosse2017almost}, and a theoretical MWPM decoder with no backlog. across both code distances and physical error rates.}
  \vspace{-25pt}
  \label{fig:comparison_backlog}
  \end{center}
\end{figure}

Notice that this accuracy tradeoff results a net resource reduction for our design over other proposed designs as shown in Figure \ref{fig:comparison_backlog}. The data backlog imposes delays into the system that decrease the logical accuracy of any decoder that incurs this backlog. As the backlog builds up, the number of required syndrome detection cycles builds up as well, resulting in a new effective logical error rate as one logical gate now requires many more syndrome detection cycles to occur. The SFQ decoder pays an accuracy price for speed, but when the backlog is taken into consideration this tradeoff results in a significant performance gain over alternative designs.


{\bf Synthesis Results and Circuit Characterization:}
Table \ref{table:synth_results} shows experimental results for the surface code decoder circuit presented in this paper using the aforementioned ERSFQ library of cells described in Section~\ref{sec:sfq_cells}. The full circuit demonstrates a cycle latency of 163 ps, and an area and power footprint of $1.28$ mm$^2$ and 13.1 $\mu$W, respectively. The full decoder is comprised of a mesh of these circuit modules, requiring a single module per individual qubit. This means that for systems of code distance $9$ comprised of 289 qubits, the decoder required will be of size $369.72$ mm$^2$ and will dissipate $3.78$ mW of power. Typical dilution refrigerators are capable of cooling up to $1-2$ Watts of power in the 4-Kelvin temperature region~\cite{hornibrook2015cryogenic}, enabling the co-location of a decoder mesh of size $~87\times 87$, which would protect a single qubit of code distance $d=44$, or $~100$ qubits of code distance $d=5$. These values are estimations given modern day SFQ and cryogenic dilution refrigerator technology, much of which is subject to change in the future. 


{\bf Execution Time Evaluation:}
The most important characteristic that the SFQ decoder aims to optimize is real-time execution speed. Previous works have described the syndrome generation time to be between $160-800$ ns for superconducting devices that we are focusing on in this study. \cite{ghosh2012surface,tannu2017taming}. 

In practice the time to solution is much lower than the upper bound of O$(n)$ on the greedy algorithm. Table~\ref{table:cycle_stats} contains the empirically observed statistics of our decoder operation. The maximum cycles to solution is well approximated by a linear scaling with a leading coefficient of $\sim 15.75$. Estimated probability distributions describing the required cycles to solution for each code distance are shown in Figure~\ref{fig:results} (c).



{\bf Comparison to existing approximation techniques:}
Trading the accuracy for decoding speed has been utilized in prior work. Union-find \cite{delfosse2017almost} achieves a significant speed-up over the minimum weight perfect matching algorithm, while the accuracy threshold decreases by only 0.4\%. Despite this, the union-find decoding time is still longer than the syndrome generation time ($>2X$ longer) thus exposing it to the exponential latency overhead caused by the data backlog. In contrast to prior approximation techniques, decoding time in our design is faster than syndrome generation time and thus it does not incur exponential latency overhead, enabling a practical implementation of error-correcting quantum machines.

{\bf Effect on SQV:}
The net effect of our design is to expand the SQV achievable by near-term machines. An example of this is a small $1,000$ physical qubit system characterized by error rates of $10^{-5}$, a machine that is an extension of one that has been conjectured will exist in the near future \cite{bishop2017quantum}. By utilizing the scaling equation described in Section \ref{sec:scaling}, we see that we can push the logical error rates achievable on this device to $2.94\times10^{-9}$ at a code distance $d=3$. This would allow for $78$ logical qubits to be packed into the device, expanding SQV from $10^5 \rightarrow \frac{1}{78\times(2.94 \times 10^9)} \approx 3.4 \times 10^8$, increasing by a factor of $3402$. This can be pushed farther by going to the small qubit count limit, constructing a machine of 40 logical qubits each of code distance 5 with logical error rate $8.96\times 10^{-10}$, yielding SQV of $\approx 1.12\times 10^9$, an increase of $11,163$. These effects are captured in Figure~\ref{fig:sqv}. Not all applications benefit from these expansions in the same fashion, but our techniques allow for machines to be used in ways that are tailored to individual applications, and enable much more computation to be performed on the same machine. 


\section{Conclusion}\label{sec:conclusion}

In the design of near-term quantum computers, it is vital to enable the machines to perform as much computation as possible. By taking inspiration from quantum error correction, we have designed an ``Approximate Quantum Error Correction" error mitigation technique that expands the ``Simple Quantum Volume" of near-term machines by factors between 3,402 and 11,163. Our design focuses on the construction of an approximate surface code decoder that can rapidly decode error syndrome, at the cost of accuracy.

Using SFQ synthesis tools, we show that the area and power are within the typical cryogenic cooling system budget. In addition, our accelerator is based on a modular, scalable architecture that uses one decoder module per each qubit.  Most importantly, our decoder constructs solutions in real-time, requiring a maximum of $\sim$20 ns to compute the solution in simulation.
This allows our decoding accelerator to achieve $10x$ smaller code distance when compared to offline decoders when accounting for decoding backlog.  Thus, it is a technique that can effectively boost the Simple Quantum Volume of near-term machines.



\bibliographystyle{IEEEtranS}
\bibliography{refs}

\end{document}